\documentclass[usenatbib]{mnras}

\usepackage{float}

\usepackage[T1]{fontenc}

\DeclareRobustCommand{\VAN}[3]{#2}
\let\VANthebibliography\thebibliography
\def\thebibliography{\DeclareRobustCommand{\VAN}[3]{##3}\VANthebibliography}

\usepackage{graphicx}	
\usepackage{amsmath}	
\usepackage{amssymb}	

\usepackage{newtxtext,newtxmath}

\title[LAEs as a probe of EoR morphology]{Lyman Alpha Emitters and the 21cm Power Spectrum as Probes of Density-Ionization Correlation in the Epoch of Reionization }

\author[Pagano \& Liu]{
Michael Pagano,$^{1}$\thanks{E-mail: michael.pagano@mail.mcgill.ca}
Adrian Liu$^{1}$\thanks{E-mail: acliu@physics.mcgill.ca}
\\
$^{1}$Department of Physics and McGill Space Institute, McGill University, Montreal, QC, Canada H3A 2T8\\
}

\date{Submitted May 21st, 2020}

\begin{document}
\label{firstpage}
\pagerange{\pageref{firstpage}--\pageref{lastpage}}
\maketitle
\begin{abstract}
Due to the large cross section of Ly$\alpha$ photons with hydrogen, Lyman Alpha Emitters (LAEs) are sensitive to the presence of neutral hydrogen in the intergalactic medium (IGM) during the Epoch of Reionization (EoR): the period in the Universe's history where neutral hydrogen in the IGM is ionized. The correlation of the ionized regions in the IGM with respect to the underlying intrinsic LAEs has a pronounced effect on the number of observed LAEs and their apparent clustering. As a result, observations of LAEs during the EoR can be used as a probe of the EoR morphology.  Here we build on previous works where we parametrize the density-ionisation correlation during the EoR, and study how the observed number density and angular correlation function (ACF) of LAEs depends on this parametrization. Using Subaru measurements of the number density of LAEs and their ACF at  z = 6.6, we place constraints on the EoR morphology. We find that measurements of LAEs at z = 6.6 alone cannot distinguish between different density-ionization models at $68\%$ credibility. However, adding information regarding the number density, and ACF, of LAEs at $z = 6.6$ to 21cm power spectrum measurements using the hydrogen Epoch of Reionization Array (HERA) at the midpoint of reionization can rule out uncorrelated and outside-in reionization at $99\%$ credibility. 

\end{abstract}




\section{Introduction}

The Epoch of Reionization (EoR) is the period in our Universe's history where the neutral hydrogen making up the intergalactic medium (IGM) is systematically ionized. The timing, duration and morphology of the EoR is still poorly understood. In order to place constraints on these quantities, a number of observational probes have been proposed. One such probe uses the hyperfine transition of hydrogen, where a $\textrm{21}$cm wavelength photon is absorbed or emitted as electrons flip their spin relative to their nucleus \citep{Furlanetto2006Review,MoralesWyitheReview,PritchardLoeb2012,LoebFurlanetto2013,LiuShawReview2020}. The advantage of using this line as a probe of the EoR is that primordial hydrogen is abundant in the early Universe and measuring the redshifting of this photon enables a three dimensional mapping of the neutral hydrogen. The photon is part of the radio spectrum and is measured in contrast to the Cosmic Microwave Background (CMB). The contrasting temperature between the photon and the CMB is referred to as a differential brightness temperature and is given by 
\begin{eqnarray}
\label{eq:dTb}
 \delta T_b(\mathbf{r}, z) \!\! &\approx& \!\!\! (27\,\textrm{mK}) \left(\frac{T_s(\mathbf{r}, z) - T_\gamma(z)}{T_s(\mathbf{r}, z)}\right)\left[1-x_{\rm HII}(\mathbf{r}, z)\right]\left[1+\delta(\mathbf{r}, z) \right] \nonumber\\
&\phantom{\times} &\times \left [\frac{H(z)/(1+z)}{dv_r/dr}\right ] \left(\frac{1+z}{10}\frac{0.15}{\Omega_mh^2}\right)^{1/2} \left(\frac{\Omega_b h^2}{0.023}\right),
\end{eqnarray}
where $\mathbf{r}$ is the position vector and $z$ is the redshift. The local ionization fraction and overdensities are given by $x_{\rm HII}(\mathbf{r}, z)$ and $\delta (\mathbf{r}, z)$ while $H(z)$ is the Hubble parameter and $dv_r/dr$ is the line of sight velocity gradient, $\Omega_b$ and $\Omega_m$ are the normalized baryon and matter densities and $h$ is the normalized Hubble parameter. The temperature $T_s(\mathbf{r}, z)$ is the spin temperature of the hydrogen atoms which describes the relative number hydrogen atoms in their excited Hyperfine states versus ground states. The CMB temperature is given by $T_\gamma$. 
The product of $x_{\rm HII}(\mathbf{r}, z)$ and $\delta (\mathbf{r}, z)$ depends on the morphology of reionization. The way these fields couple to one another in configuration space is referred to as the density-ionization correlation. In general, there are two extreme ways in which these two fields can correlate. The first is having overdense regions in $\delta$ correspond to ionized regions of  $x_{\rm {HII}}$, conversely, underdense regions in $\delta$ match neutral regions in the ionization field. This is the inside-out reionization morphology. In this model, ionization bubbles grow around overdense regions of $\delta$ until adjacent bubbles coalesce and the IGM is fully ionized \citep{FurlanettoOhPercolation2016}. The second extreme morphology is outside-in reionization. In this model, overdense regions in $\delta$ correspond with neutral regions of hydrogen. Conversely, underdense regions in $\delta$ correspond to ionized regions in $x_{\rm {HII}}$. In this scenario, the underdense regions are ionized first and the overdense regions ionized last \citep{Miralda2000,Jordan1}. We refer to inside-out reionization as having positively correlated statistics between $\delta$ and $x_{\rm{HII}}$ in contrast to outside-in reionization scenarios which has negatively correlated statistics between the ionization and density fields. The type of correlation between these fields is indicative of a particular reionization morphology.  Previous works have studied how these different models affect the brightness temperature $\delta T_b$ as well as statistical quantities that depend on it \citep{WatPrit, Binnie}. The type of correlation between these fields need not be binary, and can vary in position and as a function of redshift resulting in correlations between $\delta$ and $x_{\rm{HII}}$ with statistical combinations of inside-out and outside-in \citep{FurlanettoOhCombination2005,MadauHaardt2015}. A method to parametrize the correlation between these fields has previously been proposed by \cite{me!}.

The way these two fields correlate also has consequences for other probes of the EoR. Another such probe is the measurement of Lyman alpha flux from emitting sources during the EoR. Due to the large cross-section of Lyman alpha photons with hydrogen, even traces of neutral hydrogen left in the IGM during reionization can significantly reduce the observed Lyman alpha flux.  As a result, the number of observed LAEs is expected to significantly decrease as we observe higher redshifts, making observation of LAEs a useful probe of reionization \citep{McQuinnLAE07, GreatProbeDjistra, GreatProbeHaimanSaans, GreatProbeMaholtraRhoads, GreatProbeSimulationVerhamme, GreatProbeSantos, MasonTheUniverseIsReionizing}. Due to the sensitivity of intervening hydrogen along the line of sight, the observed number density as well as the clustering of the LAEs is influenced by the morphology of the ionization field \citep{AnneTopology, KakichiLAEMorphProbe, BoltonHaehelt}. As a result, we can make deductions about the ionization state of the IGM through measurement of the statistics of LAEs during the EoR. However in order study how the morphology of the EoR affects the flux and number count of LAEs, we must model the intrinsic luminosity of the LAEs as they leave their host halos. Assessing the intrinsic luminosity of LAEs has been difficult because the line profile of Ly$\alpha$ flux depends on the highly uncertain dynamics of the Inter Stellar Medium (ISM). As a result, the intrinsic luminosity of LAEs are subject to many uncertainties, which has led previous studies to adopt models that capture the physical range of intrinsic scenarios \citep{ChungReview, MarkDLyaReview}. 

In this paper we build on our previous work, where we parametrized the density-ionization correlations in the EoR, and study the dependence of LAE statistics on this parametrization. In order to study how the EoR morphology affects the intrinsic statistics of LAEs, we adopt a model for the intrinsic luminosity of LAEs as they leave their host haloes, which was introduced in \citealt{SobacchiAndrei, EliVisbal}. This model allows for flexibility in the physical range of intrinsic luminosities of LAEs. Using existing measurements of LAE statistics at redshift $z = 6.6$ from the Subaru Survey, we place constraints on this parametrization and therefore the EoR morphology. We then forecast the type of constraints that can be placed on the EoR morphology by combining the HSC Subaru experiment at $z = 6.6$ and a  measurement of the 21cm power spectrum using the hydrogen Epoch of Reionization Array (HERA). Since such a hypothetical 21cm measurement does not yet exist, we simulate such a measurement assuming the HERA instrument is running at its forecasted sensitivities. In these forecasts, we also include the astrophysical parameters
that have been used to model EoR physics, exploring any new degeneracies that arise from the inclusion of arbitrary density-ionization correlations.


This paper is structured as follows. In Section \ref{sec:SimulationParams} we describe the parameters used in our simulation as well as introduce our parametrization of the ionization-density correlation. At the end of Section \ref{sec:SimulationParams} we describe how the 21cm power spectrum depends on this parametrization. In Section \ref{sec:LAEmodels} we describe our model for the intrinsic luminosity of the LAEs as well as their optical depth through the IGM. In Section \ref{sec:Observables} we describe the statistical properties of LAEs and study how they vary as a function of the density-ionization correlation. In Section \ref{sec:forecasts} we describe the Subaru Survey of LAEs at $z = 6.6$ which we use to place constraints on our parametrization. Since measurements of the 21cm power spectrum do not yet exist, we also introduce our fiducial 21cm instrument (HERA) in order to forecast the constraints that can be placed on our parametrization using hypothetical measurements of the 21cm power spectrum and LAEs. Finally in Section \ref{sec:MCMCresults} we present our results and conclude in Section \ref{sec:Conclusion}. Throughout this work we set the $\Lambda$CDM parameters to $\sigma_8 = 0.81$, $\Omega_m = 0.31$, $\Omega_b = 0.048$, $h = 0.68$, consistent with Planck 2015 results \citep{Planck}


\section{EoR Simulation}
\label{sec:SimulationParams}
To generate temperature fields, halo fields, density and ionization boxes representative of different EoR models we use the \texttt{21cmFAST} package \citep{21cmFAST}. Density fields are obtained through the Zeldovich approximation while the excursion set formalism of \cite{FZH} is employed to generate the ionization and halo boxes. Using the evolved density fields, halo boxes are generated by computing the collapsed fraction of matter, larger than a threshold halo mass. The halo boxes are generated on high resolutions grids of $800^3$ voxels corresponding to a comoving side length of $200\,\textrm{Mpc}$ while the density and ionization fields use coarser boxes of $200^3$ voxels corresponding to the same comoving side length. For further details about how \texttt{21cmFAST} generates reionization models see \cite{inhomoreco}, \cite{21cmFAST}. 

We generate different EoR scenarios by varying a number of adjustable parameters whose goal is to capture variations in the detailed astrophysics of reionization. We maximize the physical range of EoR scenarios by adjusting the parameters $M_{\rm turn}$, $R_{\textrm{mfp}}$ and $\zeta$. Physically, the turnover mass $M_{\rm turn}$ determines the mass of a halo at which star formation is efficient. Values of $M_{\rm turn} \simeq 5\times 10^8 M_\odot$ correspond to a virial temperature of $T_{\textrm vir} \simeq 10^4$. Halo masses below $M_{\rm turn}$ have exponential suppression in star formation. Roughly, this sets the mass scale for the ionizing sources. The unitless astrophysical parameter $\zeta$ determines the ionizing efficiency of the sources. A large value of $\zeta$ will imply more ionizing photons per stellar baryon while a smaller ionizing efficiency will entail less ionizing photons are emitted for each ionizing source. The cutoff-radius $R_{\textrm{mfp}}$ sets the maximum size of the ionized bubbles. Recently, \cite{SteveRMFP} have developed a more realistic implementation of a cutoff-radius, which better suppresses large scale structure as compared to the standard treatment of $R_{\textrm{mfp}}$. This implementation involves modifying the excursion set conditions of \cite{FZH} to account for the excess photons required to ionize a region. We leave this for future work. Variation of these parameters affect the timing and duration of reionization and have been studied in previous studies \citep{AdrianMCMC, KernEmulator2017, AdrianParsonsParams, EwallWiceForecast2016, Park}. We use these parameters to generate a wide variety of EoR models that bracket physical scenarios. However these parameters operate under a predominately inside-out formalism in which the density field $\delta$ is positively correlated with the ionization field $x_{\rm {HII}}$ and therefore do not capture the different density-ionization correlations indicative of different EoR morphologies. In order to extend the physical scenarios bracketed by the astrophysical parameters to outside-in morphologies, we need to modify the simulation. This procedure was the focus of previous work in \cite{WatPrit} and \cite{me!}. We briefly reproduce it here.


\subsection{Extending 21cmFAST to outside-in}
\label{sec:outsidein}
In order to produce outside-in reionization morphologies we require that the temperature field $\delta T_b$ be made from density field and ionization fields which are negatively-correlated. To do this, we flip the sign of the density field at its high-redshift initial conditions \citep{PontzenPairedSimulations2016}. Once the resulting overdensity field has evolved and undergone non-linear evolution, overdense regions will be inverted to underdense regions while underdense regions will be inverted to overdense regions. The resulting ionization field will be inverted with respect to the original, non-sign flipped field. If the inverted ionization field is paired with its original, non-sign flipped, density field in Equation \eqref{eq:dTb}, then the resulting temperature field will contain the negatively correlated density-ionization statistics. The underdense regions in the temperature field are now coupled to ionized bubbles while overdense regions are coupled to neutral regions, i.e. the temperature field will contain the statistics of outside-in reionization.

\subsection{$\beta$ Parametrization}
\label{sec:beta}
We can produce EoR scenarios where the density and ionization field are correlated by arbitrary amounts. To do this we draw a random phase $\phi $ from a Gaussian of standard deviation $\sigma$, and phase shift each Fourier mode of the Fourier transformed density field $\widetilde{\delta}$, by $\phi$. When returning the overdensity box to configuration space, overdense and underdense regions in $\delta$ will have shifted from their original positions, decorrelating the density field from its original corresponding ionized fraction box. We apply this procedure to the density field at high redshift, i.e. at the initial conditions, before the density field undergoes non-linear evolution. 

\begin{figure*}
    \includegraphics[width=0.98\textwidth]{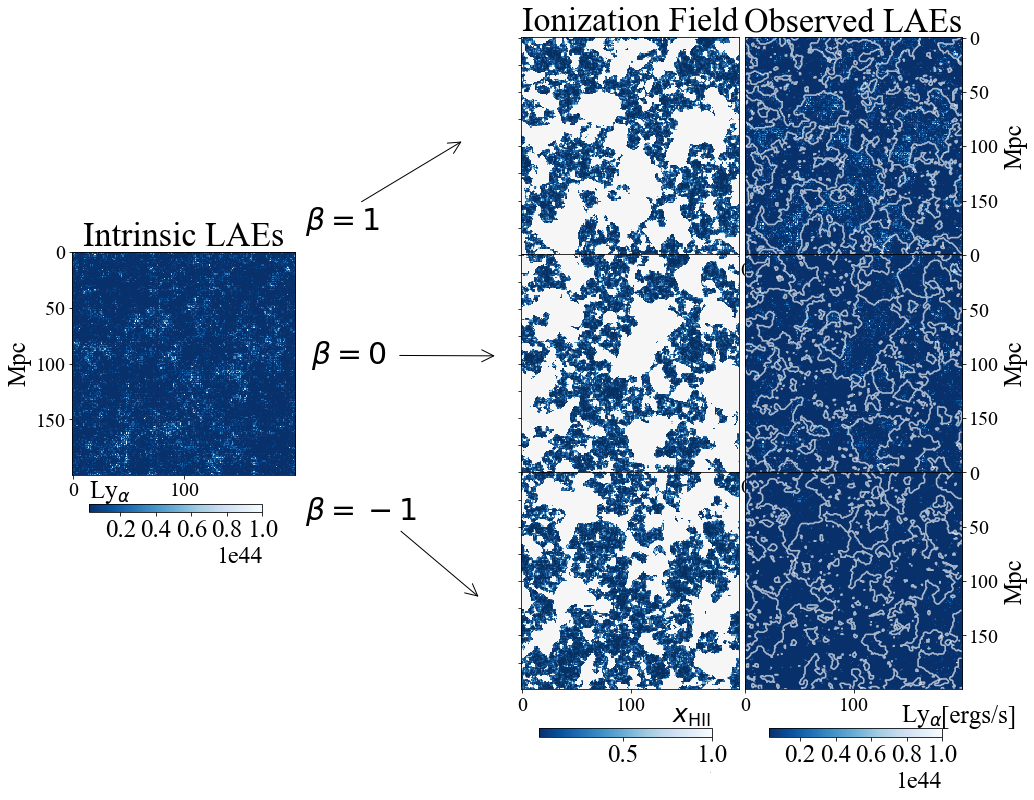}    
    \caption{Example fields demonstrating the effect that different $\beta$ correlations have on the observed LAEs. The intrinsic LAEs (left) roughly traces the underlying density field. When computing the Ly$\alpha$ optical depth using the ionization field $x_{\rm HII}$ (middle), we arrive at the observed LAEs (right). Neutral regions supress Ly$\alpha$ radiation from the observer. The ionized regions from the ionization field are superimposed as contours on the observed LAEs to emphasize the correlation between the two fields. Inside-out scenarios (with $\beta > 0$) have positively correlated density and ionization fields. Conversely, outside-in scenarios (with $\beta < 0$) have negatively correlated fields. The case $\beta = 0$ refers to the scenario where the ionization field and density field are entirely uncorrelated. These boxes are at redshift $z = 7.6$ with fiducial astrophysical parameters $\zeta_0 = 25$, M$_{\rm{turn},0} = 5\times10^8$M$_\odot$, $R_{\rm{mfp},0} = 30$Mpc.}
    \label{fig:blue_LAE_figure}
\end{figure*}

The decorrelation $\sigma$, and sign flip are folded into a single  parameter which controls the correlation between the ionization field and density field. This parameter is denoted by $\beta$ and is defined as
\begin{equation}
\label{eq:beta_def}
\beta \equiv 
\begin{cases}
\textrm{sgn}(\sigma) \left( 1 - \frac{|\sigma|}{\pi} \right) & \sigma \neq 0 \\
\pm 1 & \sigma =0
\end{cases}
\end{equation}
%
where $\textrm{sgn}(\sigma)$ is the sign of $\sigma$ which indicates whether we are decorrelating from an outside-in model ($\textrm{sgn}(\sigma)= -1$) or an inside-out model ($\textrm{sgn}(\sigma)= +1$). The case $\sigma = 0$ leads to two values of $\beta$, corresponding to the original inside-out and outside-in models. We assign these cases the values $\beta = +1$ and $\beta = -1$ respectively. The resulting parameterization can  be continuously dialled from $+1$ to $-1$ to go from a fully inside-out scenario to a fully outside-in scenario. Positive values of $\beta$ indicate scenarios where an initially correlated density and ionization field are decorrelated by $\sigma$ while a negative $\beta$ indicates a scenario where a negatively correlated density and ionization field are decorrelated by $\sigma$. We summarize the terminology of this parametization in Table \ref{tab:amps1}. We can see from Figure \ref{fig:blue_LAE_figure} that the statistics of observed LAEs is sensitive to $\beta$. In Section \ref{sec:Observables} we explore this dependence.

\begin{table}
\caption{Lexicon for physical models and their respective correlations \label{tab:amps1}}
\begin{center}
\begin{tabular}{|c|c|c|} 
\hline
$\beta$ & Moniker for Field correlations  & Physical Model \\ 
        &   $x_{\rm HII}$ $\delta $   &               \\
\hline\hline
1 &  Correlated &  Inside-out\\
\hline
  $1 > \beta > 0$ & Partially correlated & Mostly inside-out\\ 
\hline
 $0$  & Uncorrelated & Random\\ 
  \hline
 $0 < \beta < -1$ & Partially anti-correlated & Mostly outside-in\\ 
\hline
  $-1$ & Anti-correlated & Outside-in\\ 
  \hline
\end{tabular}
\end{center}
\end{table}

\subsection{Variation of $\Delta^2_{ \rm 21}(z)$ as a Function of $\beta$}
\label{sec:21cm_beta}

Our forecasts presented in Section \ref{sec:forecasts} make use of both LAE observations and measurement of the 21cm power spectrum. To gain intuition on how $\beta$ affects the 21cm power spectrum $\Delta^2_{21}$, we briefly summarize the work of \cite{me!}, which studied this in more detail. The correlation parameter $\beta$ affects $\Delta^2_{21}$, which is defined through the brightness temperature field as
\begin{equation}
\label{eq:power_spectrum}
    \Delta^2_{21}(k) \equiv \frac{k^3}{2\pi^2} \frac{ \langle | \widetilde{\delta T_b }(\mathbf{k}) |^2\rangle }{V}
\end{equation}
where $V$ is the survey volume, $\widetilde{\delta T_b}$ is the Fourier transform of the brightness temperature field (into a space defined by spatial wavevector $\mathbf{k}$), and the angular brackets indicate an average over shells of constant $k\equiv |\mathbf{k}|$.  The brightness temperature is sensitive to the inside-out versus outside-in morphology through the cross term $x_\textrm{HII} \delta$  in Equation \eqref{eq:dTb}. Consider an inside-out ($\beta = 1$) model. Decreasing $\beta$ decreases the density field's original correlation with $x_{\rm HII}$, and increases the chances that neutral regions overlap with overdense regions in $\delta$. As a result, we find increasing power on large scales as we decrease $\beta$ from $+1$ to $-1$. During the first half of reionization, the ionized bubbles are still small and so reionization has yet to make a significant imprint on the brightness temperature field. Altering the density-ionization correlation via $\beta$ thus has little effect on the power spectrum and all the $\beta$ models converge at high $z$. As one approaches a global ionization fraction of $\sim 0.5$, the ionized bubble morphology has its largest influence on the power spectrum, and thus it is there that one sees the greatest sensitivity to $\beta$. At the late stages of reionization, the IGM is increasingly ionized and $\Delta^2_{21}$ loses its sensitivity to $\beta$.

\section{LAE Models}
\label{sec:LAEmodels}

In order to infer the morphology of the EoR using Ly$\alpha$ radiation arriving at an observer, we need to model the absorption of the Ly$\alpha$ photons by the neutral hydrogen in the IGM, as well as model the intrinsic properties of the source. Our model entails two steps, assigning an intrinsic luminosity to the LAEs before the photons enter the IGM, and then computing the Lyman alpha optical depth along the line of sight, taking into account the reionization of the Universe. We first discuss the optical depth of the Lyman alpha photons and then discuss how we model the intrinsic luminosity of the LAEs.

\subsection{Ly$\alpha$ Optical Depth}
Following the approach of \cite{CharlotteGronke} and \cite{Tasitsiomi_2006}, we model the optical depth $\tau_\alpha$ for Lyman alpha photons moving through a neutral hydrogen gas cloud of number density $n_{\rm HI}(z)$ from emitted redshift $z_{\textrm e}$ to observed redshift $z_{\textrm{obs}}$ by
\begin{equation}
\label{eq:tau}
    \tau_\alpha = \int_{\textrm z_{\rm{obs}}}^{\textrm z_{e}} dz \frac{cdt}{dz} x_{\rm HI}(z) n_{\rm HI}(z) \sigma_{\alpha}(\nu, T_k) ,
\end{equation} 
where $x_{\rm HI}(z)$ is the fraction of neutral hydrogen and  $\sigma_{\alpha}(\nu, T_k)$ is cross section for Lyman alpha photons at frequency $\nu$ within a hydrogen gas cloud at temperature $T_k$. The frequency dependence of $\sigma_{\alpha}$ accounts for the redshifting of the Ly$\alpha$ photons as they move through the IGM. As the photon moves away from line center, it becomes less likely to be scattered by the intervening hydrogen. Typically the frequency dependence of $\sigma_{\alpha}$ is parametrized in terms of the dimensionless frequency $x = (\nu - \nu_{\alpha})/\Delta_{\nu_\alpha}$ which can be thought of as the ratio of frequency distance from line center to the thermal width of the line. The Ly${\alpha}$ cross section can be written as the product 
\begin{equation}
\label{eq:sigma}
     \sigma_{\alpha}(\nu, T_k) = \sigma_{\alpha_0}  \phi(x) 
\end{equation}
where $ \sigma_{\alpha_0}$ is the cross section at line center and $\phi(x)$ is a function which takes into account how $\sigma_\alpha$ varies as a function of this dimensionless frequency (as it moves along its line of sight). The cross section $\sigma_{\alpha_0}$ at line center is given by
\begin{equation}
\label{eq:sigma0}
    \sigma_{\alpha_0} = \frac{1}{\sqrt{\pi} \Delta_{\nu_\alpha}} \frac{f_\alpha \pi e^2 }{m_e c} \simeq 5.9 \times 10^{-14} \sqrt\frac{T_k}{10^4 K}
\end{equation} 
where $f_\alpha = 0.416$ is the Ly$\alpha$ oscillator strength, $e$ is the charge of the electron, $m_e$ is the mass of the electron and . Typically how $\sigma_{\alpha}$ varies along the line of sight can be broken into two regimes: (1) frequencies close to line center which we refer to as the core of the line and frequencies further away from line center which we refer to as the wing. Both these regimes can folded into the Voigt function defined by:
\begin{equation}
\label{eq:voigt}
\phi(x) = \frac{a_\nu}{\pi}\int_{-\infty}^{\infty} dy \frac{e^{-y^2}}{(y-x)^2 + a_\nu^2}
\end{equation}
where $a_\nu = 4.7\times 10^{-4} \sqrt\frac{T_k}{10^4 K}$ is the Voigt parameter. To evaluate Equation \eqref{eq:voigt}, we use the approximation made in \cite{Tasitsiomi_2006}. The cross section $\sigma_{\alpha}(\nu, T_k)$ is tightly peaked close to line center $\nu_{\alpha_0}$ and then drops rapidly as a function of $x$. The cross-over from core to wing occurs at $x \simeq 3$, which occurs on sub-grid scales in our simulation. In order to properly model the absorption of Ly$\alpha$ photons by neutral hydrogen within the core of the line, one must have sufficient resolutions of $n_{\textrm {HI}}$ and  $x_{\rm{HII}}$ to track the propagation of photons for $x < 3$ which correspond to physical scales of $\ll 1$Mpc . Since our density and ionization boxes described in Section \ref{sec:SimulationParams} have resolution of $~ 1.5$Mpc per pixel we use a weighted average of $\sigma(x)$ for values of $x \ll 3$ and then switch to wing absorption through the wing for $x > 3$ through Equation \eqref{eq:voigt}. This crossover point from core to wing has a weak dependence on the temperature $T_k$ of the gas. Depending on the particular reionization model chosen, the temperature $T_k$ of the gas has a physically motivated range of $1$K$ < T_k < 10^3$K. There are physically motivated arguments for both extreme temperature scenarios; the $10^3$K scenario is due to x-ray heating of the neutral IGM, for example. In the neutral IGM, we set the temperatures of the gas to be $T_k = 1$K which is consistent with \cite{CharlotteGronke}. Our conclusions presented in Section \ref{sec:MCMCresults} do not depend on how one chooses to model the temperature of the gas. The temperature of ionized bubbles is set at $T = 10^4$K which is consistent with a photoionised gas at the mean density \citep{ionizedtemperature}.

\begin{figure}
    \includegraphics[width=8cm]{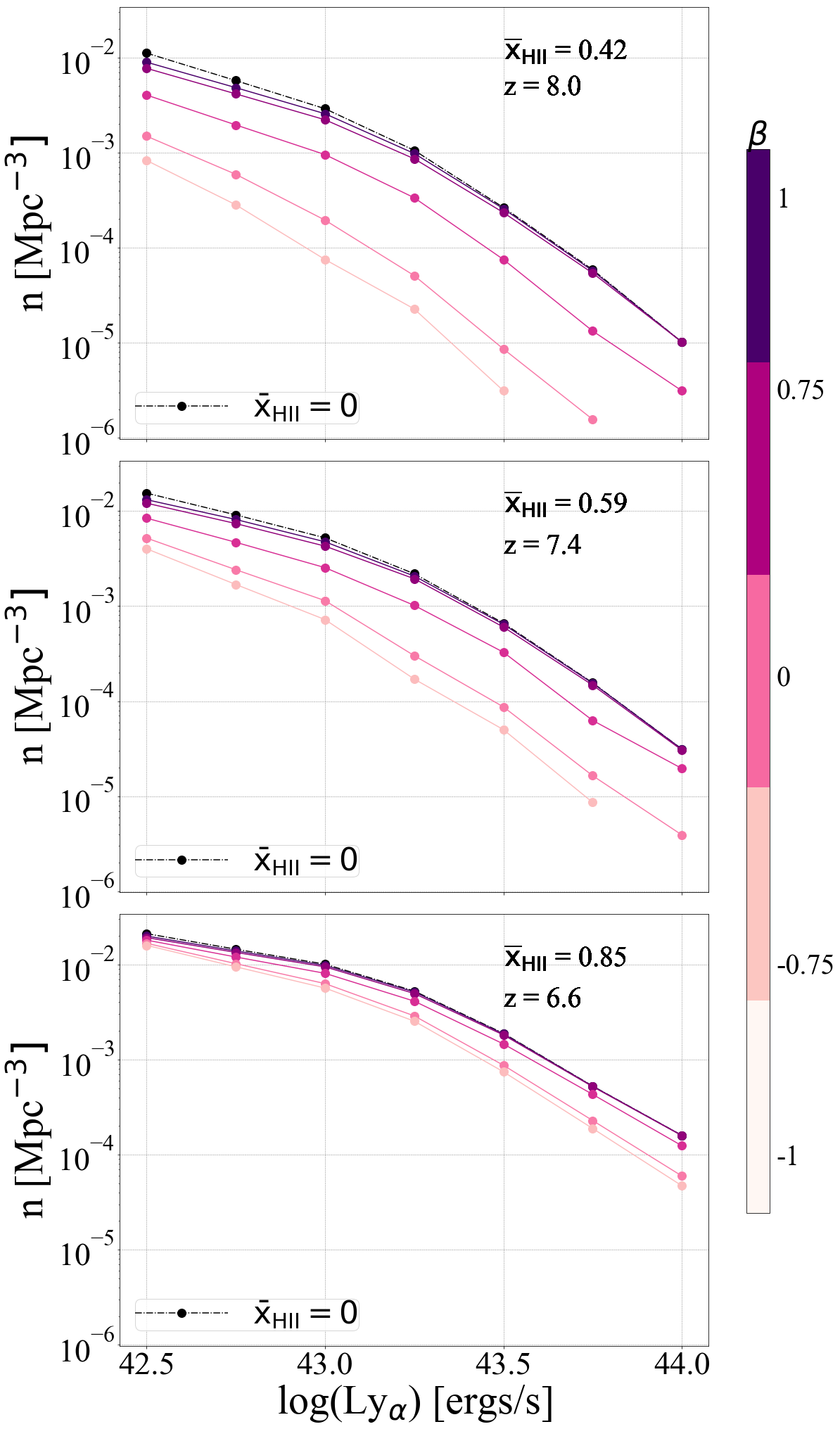}
    \caption{The Ly$\alpha$ luminosity function of LAEs as a function of $\beta$, at three different stages of reionization. In each panel, the dotted curve corresponds to the luminosity function of the intrinsic field (i.e. with $\overline{\rm{x}}_{\rm{HII}} = 0$). Notice that extreme outside-in reionization scenarios ($\beta \sim -1)$, lead to concealing the intrinsically brightest LAEs. At high redshift, where there are fewer intrinsically bright LAEs, this leads to a sharp drop-off of the Ly$\alpha$ luminosity function. In each of these curves, the detection threshold Ly$_\alpha^{\rm{min}}$ corresponds to mass M$_\alpha^{\rm{min}} \sim 10^{10}$M$_{\odot}$ with $f_{\rm{duty}} = 1$.}
    \label{fig:Luminosity_Function}
\end{figure}

\begin{figure*}
    \includegraphics[width=0.99\textwidth]{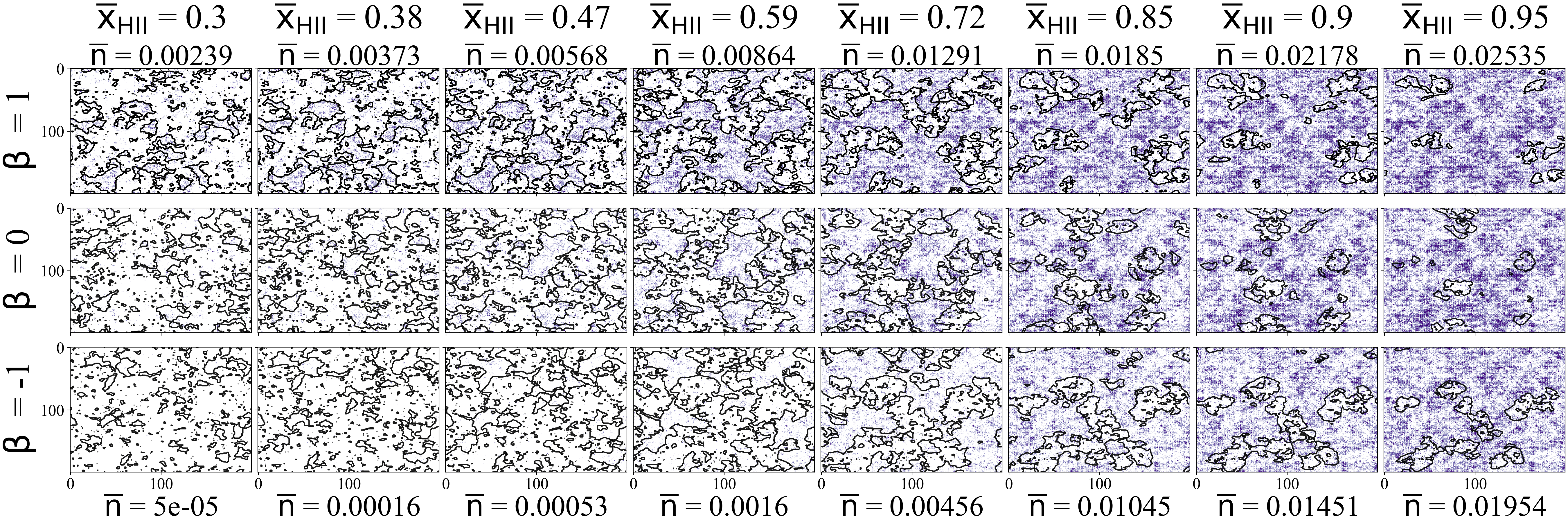}
    \caption{Evolution in the number of observed LAEs for positively correlated models ($\beta > 1$), uncorrelated models ($\beta = 0$), and negatively correlated models ($\beta = -1$) at different mean ionized fractions. The mean number density of LAEs $\overline{n}$, for the extreme $\beta = 1$ and $\beta = -1$ models are noted. Notice the rapid change in the number of LAEs for $\beta = -1$ models. As the ionized regions grow large enough to encompass the high density regions, the number of LAEs increases rapidly. The uncorrelated model has ionized regions which are random with respect to the underlying density field. Positively correlated models have the ionized regions that always correspond to the high density regions, where the intrinsic number density of LAEs is greatest. }
    \label{fig:evolution_of_ndensity}
\end{figure*}

\subsection{LAE Intrinsic Luminosity}
\label{sec:intrinsicLuminosity}

We assume that the intrinsic Ly$\alpha$ luminosity of the emitters is proportional the mass of the halo in which it resides. This model is similar to one used in \cite{SobacchiAndrei} and \cite{EliVisbal}. This method does not intend to capture the small scale radiative transfer physics in the ISM which are required to realistically model the intrinsic luminosities of LAEs. Rather, we use this framework to bracket the physical range of intrinsic clustering by providing flexibility to the high degree of uncertainty regarding the intrinsic properties of the source. We then study how the EoR morphology, imprinted through the opacity in Equation \eqref{eq:tau}, affects the luminosity, clustering and number densities of the intrinsic LAEs. The intrinsic luminosity of a Ly$\alpha$ emitter residing in a halo of mass $M_h$, before the Ly$\alpha$ flux is altered by the IGM, is given by,
\begin{equation}
    \label{eq:L_intrinisc}
    \textrm{L}_\alpha^{\textrm {int}} = L_{\alpha}^{\textrm {min}}  \left(\frac{M_h}{M_{\alpha}^{\textrm{min}}} \right)^\gamma \chi 
\end{equation}
where $\chi = 0$ or $1$ is a random variable that encodes the uncertainty whether a given halo hosts an LAE. The probability that a halo hosts an LAE, i.e. the probability that $\chi = 1$, is equal to the duty cycle $f_{\rm{duty}}$. We assume $f_{\rm{duty}}$ to be independent of its mass. The luminosity $ L_{\alpha}^{\textrm {min}}$ is the instrumental threshold for detection of an LAE, which we set to be $ L_{\alpha}^{\textrm {min}} = 2.5\times 10^{42}$ ergs/s, corresponding to the detection limit of the HSC Ultra Deep Field at redshift $z = 6.6$. The mass $M_{\alpha}^{\textrm{min}}$ is the halo mass corresponding to the detection threshold of $L_{\alpha}^{\textrm {min}}$. Haloes with masses less than $M_{\alpha}^{\textrm{min}}$ lead to intrinsic LAE luminosities that are below the detection threshold and so unobservable. The factor of $\gamma$ is the power law index which controls the inclination of the intrinsic LAE luminosity function. These parameters provide the necessary flexibility in the intrinsic Ly$\alpha$ luminosity function to bracket the physical range of the clustering signal \citep{SobacchiAndrei}. Take for example the normalization constant $M_\alpha^{\rm min}$, which shifts the intrinsic luminosity function of the LAEs left and right. Values of $M_{\alpha}^{\textrm{min}} \simeq 10^{11}$M$_\odot$ allow only the largest LAEs to be observable (i.e. above the detection threshold $L_{\alpha}^{\textrm {min}}$). This leads to a scenario where only the largest haloes contribute to the survey. In contrast, a lower $M_{\alpha}^{\textrm{min}}$ means that the smaller LAEs will also be observable by our instruments, which allow us to measure LAEs in the underdense regions of the Universe. The normalization mass $M_{\alpha}^{\textrm{min}}$ will therefore affect the statistics of the intrinsic LAEs The duty cycle $f_{\rm{duty}}$, adjusts the number of halos in the survey which shifts the intrinsic luminosity function up and down. The parameter $f_{\rm{duty}}$ affects the number densities of intrinsic LAEs but doesn't affect the clustering of the intrinsic LAEs. Finally $\gamma$ tilts the the LAE luminosity function. For example, reducing $\gamma$ from $1$ to $\gamma = 2/3$ decreases the number of intrinsically bright LAEs and increases the number of faint LAEs. Varying $\gamma$ within the range $1/2 \le \gamma \le 1$ does not significantly affect the clustering signal of the LAEs or the observed number density of LAEs (see Sections \ref{sec:n}, \ref{sec:ACF}). For the remainder of the analysis we set $\gamma = 1$ without loss of generality. A possible shortcoming of this model is that the brightest LAEs are assumed to reside in the overdense regions of $\delta$, the location of the largest halos. This might not be the case if one takes into account the UV reddening of dusty galaxies in the vicinity of larger halos. Such a scenario would cause an apparent dimming of bright LAEs hosted in large dusty halos \citep{2020JordanRecent}. We leave these broader range of scenarios for future work.

We assign an intrinsic Ly$\alpha$ luminosity to each halo in our simulation box according to Equation \eqref{eq:L_intrinisc}. The observed luminosity of the LAE after the Ly$\alpha$ photons pass through the IGM is given by
\begin{equation}
    \label{eq:L_obs}
    L_\alpha =\textrm{L}_\alpha^{\textrm {int}}e^{-\tau_{\alpha}}  ,
\end{equation}
where $e^{-\tau_{\alpha}}$ is computed by integrating Equation \eqref{eq:tau} along the line of sight for a given reionization scenario. LAEs with apparent luminosity $L_\alpha$ satisfying $L_\alpha < L_{\alpha}^{\rm{min}}$ are removed from our mock survey since they have Ly$\alpha$ luminosity below the detection threshold. To mimic the LAEs observed by the Subaru Survey, which has redshift thickness $\Delta z = 0.1$ (corresponding to $\sim 37$Mpc at $z = 6.6$), we slice our observed LAEs corresponding to the same redshift thickness, $\Delta z = 0.1$ (see Section \ref{sec:subaru}).  In the following Section we study how the morphology of the EoR affects the observability of these LAEs.



\section{Dependence of LAE Statistics on $\beta$}
\label{sec:Observables}


The observed Ly$\alpha$ flux from a LAE is very sensitive to any neutral hydrogen that lies along its line of sight. Consequently, the Ly$\alpha$ flux from LAEs which reside near neutral regions is severely attenuated. These LAEs are less likely to be observed over the detection threshold $L_{\alpha}^{\rm min}$. Conversely, LAEs coupled to regions of high $x_{\rm HII}$ (ionized bubbles) are more likely to be observed. Since the underlying intrinsic LAEs roughly trace the density field $\delta$, we can control how the intrinsic LAEs couple to $x_{\rm HII}$ by varying $\beta$, which controls the coupling between $\delta$ and $x_{\rm HII}$. In this Section, we study the effect that different EoR morphologies have on our measurements of the LAEs. In Section \ref{sec:luminosityfunction} and \ref{sec:n} we study how $\beta$ affects the LAE luminosity function and mean number density of LAEs, while in Section \ref{sec:ACF} we study how the clustering of LAEs depends on the EoR morphology.

\subsection{Dependence of the Observed LAE Luminosity Function on $\beta$}
\label{sec:luminosityfunction}
Let us consider the underlying intrinsic LAEs at redshift $z$. Recall that the value of $\beta$ changes how the ionization field couples to the intrinsic LAEs. Values of $\beta \simeq 1$ imply an inside-out reionization scenario where the ionization field is correlated with the underlying density field. In this scenario, LAEs found in overdense regions will correspond to regions of high $x_{\rm {HII}}$ which reduces the effect of the attenuation factor in Equation \eqref{eq:tau} and makes these LAEs more likely to be observed. Conversely, LAEs which reside in underdense regions of $\delta$ will correspond to regions of low $x_{\rm {HII}}$, where the presence of neutral hydrogen will obscure them.  Since the overdense regions are more likely to host the intrinsically brightest LAEs, $\beta = 1$ will tend to allow the brightest LAEs to be observed. This maximizes the amount of intrinsically bright LAEs which are observable. We can see the effect that an inside-out reionization scenario has on the observability of the LAEs in Figure \ref{fig:blue_LAE_figure}.

Figure \ref{fig:Luminosity_Function} illustrates the effect of decreasing $\beta$ on the LAE luminosity function. As we decrease $\beta$ from $\beta = 1$, ionized bubbles in $x_{\rm {HII}}$ become increasingly decorrelated from the underlying density field $\delta$, and so the placement of the ionized regions are increasingly randomized in relation to the intrinsic LAEs. This means that some overdense regions in $\delta$ will now couple to regions of low $x_{\rm {HII}}$, obscuring the LAEs which reside in that region. This leads to a decrease in the bright end of the LAE luminosity function from the purely inside-out ($\beta = 1$) scenario. As we continue to decrease $\beta$ to $\beta \simeq -1$, corresponding to an outside-in reionization scenario, ionized bubbles will be increasingly coupled to underdense regions in $\delta$. LAEs which reside in the underdense regions become more likely to be observed by experiment, while the flux from LAEs which reside in overdense regions in $\delta$ will be coupled to regions of low $x_{\rm {HII}}$, and therefore severely attenuated. Since the brightest LAEs reside in the overdense regions, we notice a sharp decrease in the bright end of the LAE luminosity function. We can see the sharp dropoff of bright LAEs for outside-in driven models in Figure \ref{fig:Luminosity_Function}. This sharp dropoff of bright LAEs for outside-in driven models becomes even more pronounced at higher redshift where there are fewer intrinsically bright LAEs. As outside-in reionization proceeds, the ionized regions grow and expose the overdense regions where the brightest LAEs reside. By the end of outside-in reionization, only the intrinsically brightest LAEs are still unobservable. Referring again to Figure \ref{fig:Luminosity_Function}, this tilts the Ly$\alpha$ luminosity function compared to extreme inside-out models.

 \begin{figure}
  \includegraphics[width=0.5\textwidth]{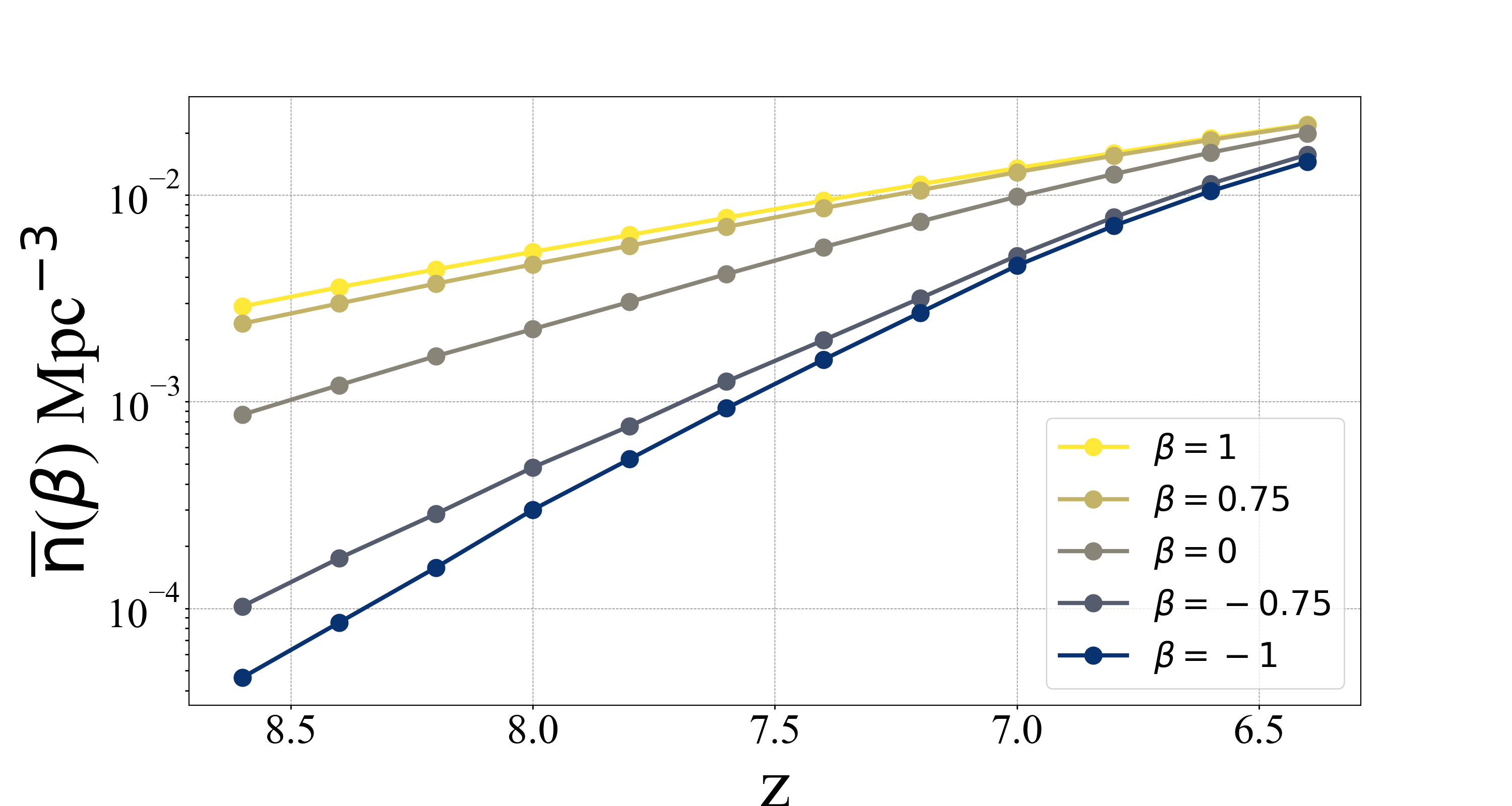}
  \caption{Evolution of the number density of LAEs for different reionization morphologies. Inside-out driven models ($\beta > 0$) initially have more observable LAEs since the intrinsically densest regions of the IGM are ionized first. As the ionized regions grow to encapsulate the underdense regions in $\delta$, the mean number density grows slowly as compared to outside-in driven models ($\beta < 0$), where the underdense regions are first to be ionized. In this scenario, the number density of LAEs increases rapidly as the intrinsically densest regions are ionized.}
  \label{fig:number_density_beta_z}
\end{figure}

\subsection{Dependence of the Mean Number Density of LAEs on $\beta$}
\label{sec:n}
From the LAE luminosity function we can extract another useful quantity, the mean number density $\overline{n}$, of the observed LAEs. This quantity represents the number of observable LAEs per unit volume and can be derived by summing over the luminosity function:
\begin{equation}
\label{eq:n}
    \overline{n} = \int_{M_\alpha^{\rm min}}^\infty n(M_h) d M_h   ,
\end{equation}
where $M_\alpha^{\rm min}$ corresponds to the detection threshold $L_\alpha^{\rm min}$. The number density of LAEs has already been constrained to be $\overline{n} = 4.1^{+0.9}_{-0.8}\times 10 ^{-4}$Mpc$^{-3}$ at $z = 6.6$ corresponding to a minimum threshold luminosity of $L_\alpha^{\rm min} = 2.5 \times 10^{42}$ergs/s   \citep{Ouchi}. The EoR morphology will influence which LAEs in the intrinsic field are observable, and so will influence the measured number density of LAEs, $n_{\rm{obs}}$. 
To see how $\beta$ influences $n_{\rm{obs}}$, note that the intrinsic LAEs trace the underlying halo field (Equation \eqref{eq:L_intrinisc}). Therefore overdense regions of $\delta$ contain many more LAEs than the underdense regions of $\delta$.  For inside-out reionization scenarios ($\beta = 1$), overdense regions in $\delta$ correspond to regions of high $x_{\rm {HII}}$ in the ionization field. The intrinsically densest regions of LAEs are most likely to be observed first. As the ionized bubbles grow, the underdense regions in $\delta$ become observable which contain statistically fewer LAEs. The measured number density of LAEs,  $n_{\rm{obs}}$,  increases slowly as reionization progresses. In Figure \ref{fig:evolution_of_ndensity} we show the LAE number field. We can see the evolution in the number of LAEs for $\beta = 1$ in the top row of Figure \ref{fig:evolution_of_ndensity}. Notice that since the densest regions are ionized first, LAEs are observable at low mean ionized fraction. Consider now the outside-in model of reionization ($\beta = -1$), where the underdense regions are ionized first. Since the underdense regions are less likely to host LAEs, then only the intrinsically sparest regions of LAEs are observable early in reionization. This may make high redshift LAEs difficult to find for outside-in models \citep{MasonHardToSeez7}. However, as reionization progresses, the ionized regions grow and overdense regions are increasingly coupled to low $x_{\rm{HII}}$ which allows the regions with the intrinsically densest regions of LAEs to be observed. From the bottom row of Figure \ref{fig:evolution_of_ndensity} we can see that there are fewer LAEs observable at low mean ionized fraction compared to $\beta = 1$ models. In Figure \ref{fig:number_density_beta_z} we can see how this translates to the redshift evolution of $\overline{n}_{\rm{obs}}$ for various $\beta$ values. Outside-in models produce scenarios where $n_{\rm {obs}}$ increases rapidly as reionization progresses. As we increase $\beta$ from its extremum, $\beta = -1$, the number density, $\overline{n}_{\rm{obs}}$, increases monotonically until it is maximized with respect to $\beta$ for inside-out reionization scenarios ($\beta = 1$). As reionization progresses, the IGM becomes increasingly ionized and the number density of LAEs become insensitive to $\beta$. All models converge to the intrinsic number density of LAEs (see Figure \ref{fig:number_density_beta_z}). In Section \ref{sec:forecasts}, we use existing constraints on $n_{\rm{obs}}$ at $z = 6.6$ to place constraints on $\beta$. In the following Section, we introduce the angular correlation function (ACF) as another statistical tool to study LAEs, which along with $n_{\rm{obs}}$, has already been constrained at $z = 6.6$.


\subsection{Dependence of the ACF on $\beta$}
\label{sec:ACF}
The probability of finding a pair of LAEs at a distance $R$ from one another is 
\begin{equation}
\label{eq:Poisson}
    dP_{\rm{ 1 2}} = \overline{n}^2 \left [ 1 + \xi(R) \right ] dV_1 dV_2
\end{equation}

\begin{figure*}
    \includegraphics[width=0.9\textwidth]{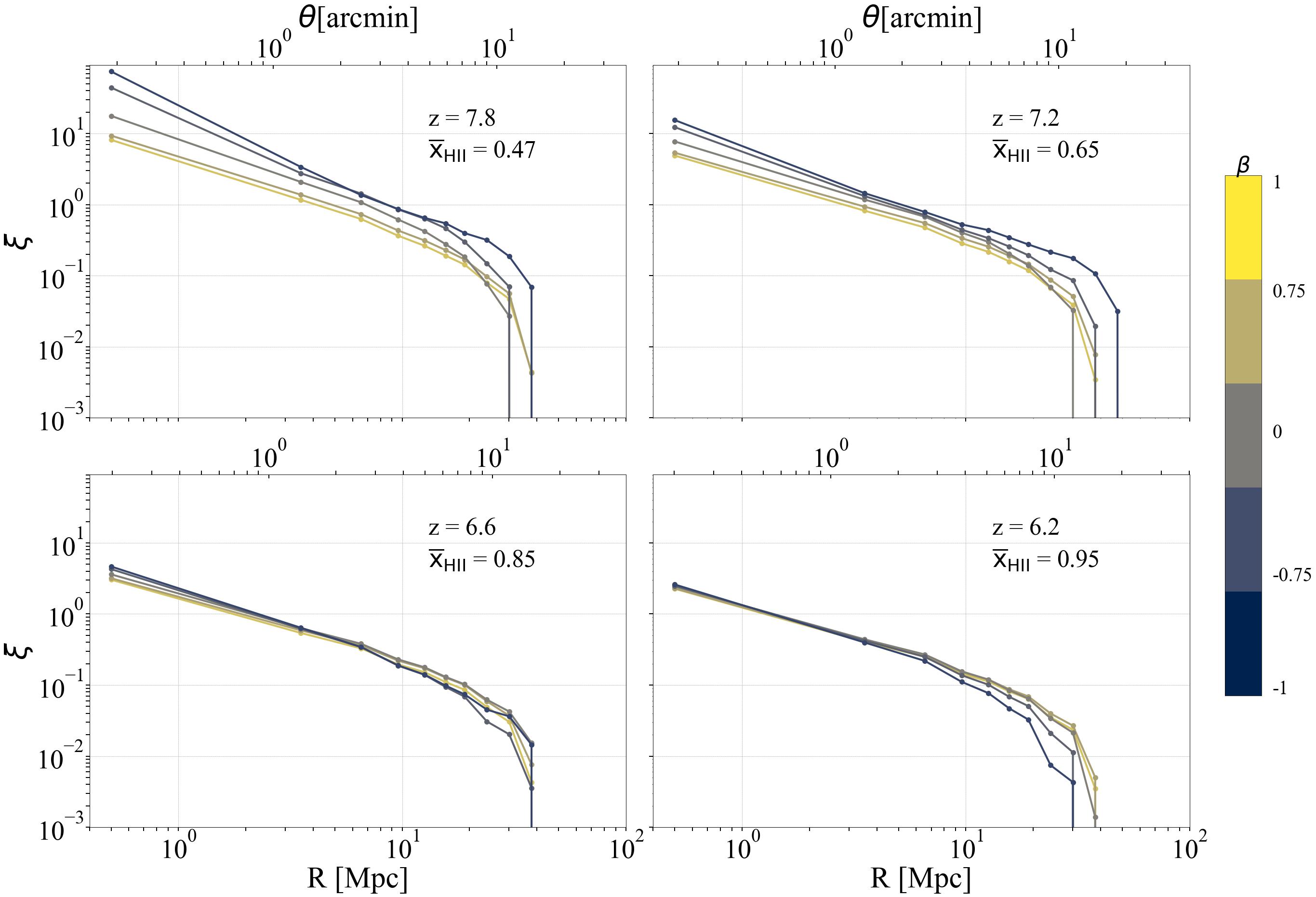}
    \caption{ACF of various $\beta$ scenarios at different stages of reionization. Outside-in driven reionization scenarios ($\beta < 0$) produce the largest clustering of LAEs. The contrast between the different $\beta$ models is greatest at high redshifts where the ionization of the IGM has the largest imprint on LAE observability. As reionization proceeds it becomes more difficult to distinguish between these scenarios. We use fiducial parameters $\zeta_0 = 25$, $M_{\textrm{turn}, 0} = 5\times 10^8 M_{\odot}$, $R_{\textrm{mfp},0} = 30\,\textrm{Mpc}$, and $\beta_0 = 0.936$, $f_{\rm{duty}} = 1$ and M$_\alpha^{\rm{min}} = 10^{10}$M$_\odot$.}
    \label{fig:ACF}
\end{figure*}

\begin{figure}
    \includegraphics[width=0.4\textwidth]{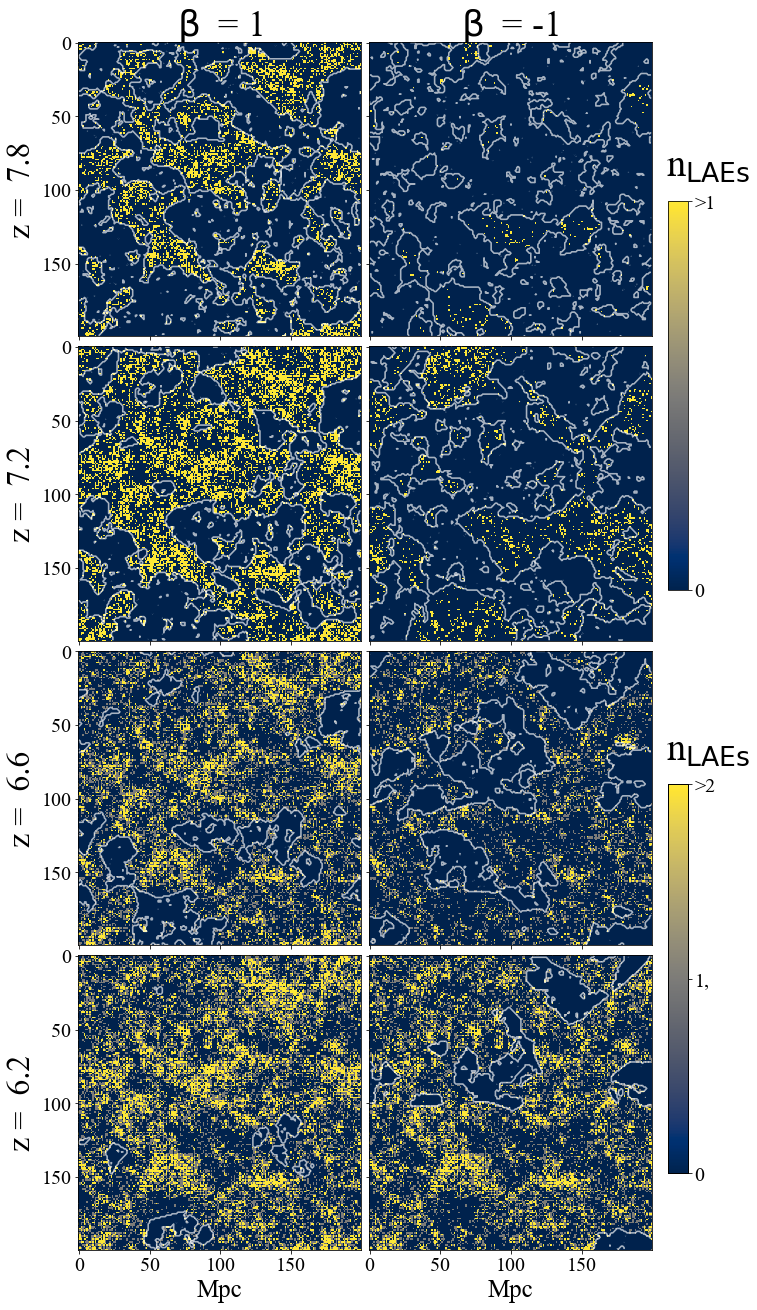}
    \caption{The number density of observed LAEs for different $\beta$ models and at different redshifts. For inside-out models at high redshifts (top two rows), LAEs are spread across the entire ionized regions, while outside-in models produce fewer LAEs which are spread across a limited volume. This serves to increase the clustering of LAEs for $\beta = -1$ models. The scale is binary to accentuate the difference in the spatial distributions of LAEs between the two models. Later in reionization (bottom two rows), the LAEs of both inside-out and outside-in models are spread over similar ionized volumes. However the number density of LAEs within these ionized volumes is larger for inside-out models which serves to increase the clustering of LAEs for $\beta = -1$ models compared to $\beta  = +1$. We use fiducial parameters $\zeta_0 = 25$, $M_{\textrm{turn}, 0} = 5\times 10^8 M_{\odot}$, $R_{\textrm{mfp},0} = 30\,\textrm{Mpc}$, and $\beta_0 = 0.936$, $f_{\rm{duty}} = 1$ and M$_\alpha^{\rm{min}} = 10^{10}$M$_\odot$. }
    \label{fig:clustering}
\end{figure}

\noindent where $\overline{n}$ is the mean number density of LAEs, $dV_1$, and $dV_2$, are volume elements of the survey in the vicinity of each LAE, and $\xi(R)$ is the two point correlation function. The two point angular correlation function is the excess probability as compared to a random Poisson distribution, that two LAEs be located a distance $R$ from one another. To compute $\xi(R)$, we first form the overdensity $\delta_n$ :
\begin{equation}
\label{eq:overdensity_n}
\delta_n(\mathbf{x},z) = \frac{n_{\rm {LAE}}(\mathbf{x},z)}{\overline{n}} -1  ,
\end{equation}
where $\overline{n}$ is the mean number of LAEs at redshift $z$ and $n_{\rm {LAE}}(\mathbf{x},z)$ is the number density field which describes the fluctuations in the number of LAEs about the mean (according to position $\mathbf{x}$ and redshift $z$). We can compute $\xi(R)$ directly from the Fourier transformed overdensity field $\widetilde{\delta_n}$, 
\begin{equation}
    \label{eq:ACFfromn}
    \xi(R) = \left < \int d^3\mathbf{k} e^{i \mathbf{k}\cdot \mathbf{x}}  | \widetilde{\delta_n}(\mathbf{k}) |^2 \right>_{\mathbf{x} \in R}   .
\end{equation}
where the angular brackets indicate a spatial average. Physically, we interpret the correlation function as the clustering of the LAEs at separation $R$. Larger values of $\xi$ imply more clustering of LAEs. The two point correlation function $\xi$ of LAEs can be expressed as a function of angular separation $\theta$ on the sky. This is the angular correlation function (ACF), denoted by $\xi(\theta)$. Note that in a LAE survey, one actually measures $\xi(\theta)$. However, to build intuition in our theory interpretation, we use $\xi(R)$ to study the clustering of LAEs. For small angular separations and thin layers in $\Delta z$, we can simply convert $\xi(R)$ to $\xi(\theta)$ using $D_c$ which is a conversion factor from $ \theta$ to transverse comoving distance $R$, and is given by
\begin{equation}
\label{eq:X_convert}
    D_c  \equiv \frac{R}{\theta} = \frac{c}{H_0}\int^z_0 \frac{dz'}{E(z')}
\end{equation}
with $c$ the speed of light, $H_0$ the Hubble parameter today, $E(z) \equiv \sqrt{\Omega_m(1+z)^3 + \Omega_\Lambda}$ and $\Omega_\Lambda$ the normalized dark energy density \citep{Simon}. For the remainder of this paper we work with $\xi(R)$, recognizing that one can easily convert $\xi(R)$ to $\xi(\theta)$ using Equation \eqref{eq:X_convert} under the Limber approximation \citep{Simon}.

The ACFs of the intrinsic LAEs under different density-ionisation correlation scenarios are shown in Figure \ref{fig:ACF}. We split our discussion of the $\beta$ dependent ACF into two different redshift regimes: higher redshifts (Figure \ref{fig:ACF} top two rows), and lower redshifts (Figure \ref{fig:ACF} bottom two rows). Consider first the higher redshift regime, outside-in scenarios ($\beta = -1$), lead to more clustering as compared to inside-out models ($\beta = 1$). To see why this is, consider LAEs in the ionized regions of an outside-in ($\beta = -1$) scenario, where the ionized regions correspond to underdense regions in $\delta$. We can see from the top two rows in Figure \ref{fig:clustering}, the LAEs are rare and occupy an only fairly limited portion of the ionized volume. Due to the limited volume that they occupy, any LAEs observed in $\beta = -1$ models tend to be clustered together. This produces a strong clustering signal for $\beta = -1$ models. Now let us consider LAEs in the ionized regions of inside-out models. There is a significant increase in the number of LAEs compared to outside-in models. Referring again to the top two rows in Figure \ref{fig:clustering}, the excess LAEs in $\beta = 1$ models are spread over the entire volume of the ionized region. This reduces the excess probability $\xi(R)$ in Equation \eqref{eq:Poisson} of finding LAEs separated by distances smaller than the bubble size, leading to a decrease in clustering compared to outside-in driven models. As we decrease $\beta$ from $\beta = 1$ to $\beta = -1$, we find that there is an increase in the clustering signal. This conclusion is also true when the extreme $\beta$ models are constrained to have a fixed number density of LAEs. We can study this scenario by tuning the duty cycle $f_{\rm{duty}}$ of both models, so that they have the same number density of LAEs. This entails decreasing f$_{\rm{duty}}$ for $\beta = 1$ models, such that the number of LAEs within the ionized regions are the same as for $\beta = -1$ models. In this scenario the LAEs for the inside-out scenario are still spread over a larger volume as compared to the LAEs within the ionized regions of outside-in models, leading to the same conclusions as above. 

\begin{figure}
\includegraphics[width=0.47\textwidth]{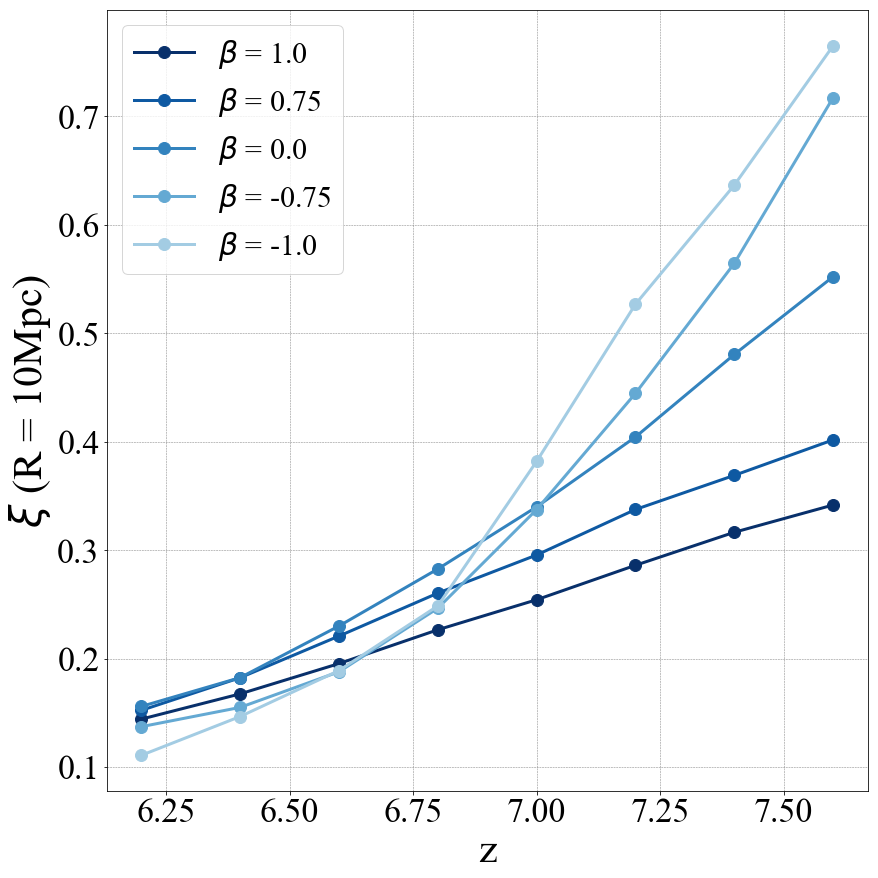}
    \caption{Angular correlation function as a function of $z$ for various $\beta$ models at separations of 10Mpc. Note the non-monotonic behaviour of $\beta$. Early in reionization, the LAEs of inside-out models are spread over larger volumes within the ionized regions as compared to outside-in models, where the LAEs tend to be more localized. This produces a stronger clustering signal for outside-in driven models. This behaviour is inverted later in reionization where there are enough intrinsic LAEs to fill the entire volume of ionized regions for outside-in models. }
    \label{fig:omega_10Mpc}
\end{figure}

Our above conclusions are essentially due to LAEs being spread over a larger ionized volume for inside-out models, while outside-in models produce localized fluctuations of LAEs within the ionized regions. Let us now focus on the behavior of $\xi$ at lower redshifts where this is no longer true. At the end of reionization, more structures have collapsed to form haloes, and so the number of LAEs in the underdense regions of the intrinsic field dramatically increases. Referring to the bottom two rows in Figure \ref{fig:clustering}, the ionized regions of outside-in reionization maps now contain LAEs that are spread over the entire ionized volume. Consider LAEs separated by $R$ within these ionized volumes, where $R$ is much smaller than the typical ionized bubble. Since there are more LAEs within the ionized volumes of inside-out driven models compared to $\beta = -1$ models, there are a larger fraction of LAEs separated within $R$ increasing the clustering compared to outside-in driven models. Equivalently, the ionized volumes of inside-out regions allow us to observe the highest mass halos, which tend to be the most biased traces tracers of the density field, i.e. the most clustered. In Figure \ref{fig:omega_10Mpc}, we can see the non-monotonic behaviour of $\beta$ in $\xi$ at separations of $R = 10$Mpc, a length scale entirely contained within the ionized regions at these redshifts. At low redshifts, the outside-in driven models $(\beta <0)$, produce the smallest clustering of LAEs. As reionization continues to proceed, the ionized regions grow, exposing both overdense and underdense regions, which narrows the contrast of $\xi(R)$ between the extreme models, $\beta = 1$ and $\beta = 1$. Finally as the IGM is entirely ionized, the clustering signatures of the extreme models become indistinguishable and $\xi(R)$ is no longer sensitive to $\beta$. 

In Section \ref{sec:MCMCresults}, we shall see that at redshift $z = 6.6$, the differences in $\xi(R)$ between extreme models are not significant enough to distinguish between them using existing data. Since the astrophysical parameters $M_{\rm {turn}}$,  $R_{\rm {mfp}}$ and $\zeta$, affect the size of the ionized regions at each $z$, they will also influence the observed clustering of and number density of LAEs. In the next section we use existing measurements of $\xi(R)$ and $\overline{n}$ to place constraints on these parameters as well as $\beta$.

\section{Forecasts and Constraints}
\label{sec:forecasts}
In Section \ref{sec:beta} we developed a framework where we can track the density-ionization correlations during the EoR. In Section \ref{sec:Observables}, we developed an intuition on how the observed clustering, number density and luminosity function of the LAEs depend on this correlation. In this section, we use measurements of the number density and clustering of LAEs at $z = 6.6$ made by the Subaru survey to place constraints on the correlation parameter $\beta$, as well as the other EoR parameters. Further, we forecast the type of constraints that can be placed on $\beta$ using a joint measurement of LAE and the 21cm power spectrum using HERA. We discuss the HERA instrument as well as its fiducial instrument parameters. We present the results of these forecasts in Section \ref{sec:MCMCresults}.

\subsection{Subaru Survey}
\label{sec:subaru}
To place constraints on our EoR parameters including $\beta$, we use measurements of $\overline{n}$ and $\xi(\theta)$ at redshift $6.6$ from the Subaru/XMM-Newton Deep Survey (SXDS) survey \citep{Ouchi}. Here we summarize the analysis done by \cite{Ouchi} in computing $\overline{n}$ and $\xi(\theta)$. Measurement of $\overline{n}$ is made indirectly by first measuring the $Ly\alpha$ luminosity function using the $Ly\alpha$ line profiles based on 207 $Ly\alpha$ emitters at z = 6.6 on the $1$-deg$^2$ sky, which have redshift uncertainty $\Delta z \simeq 0.1$. Using these measurements, the constraints are placed on the $Ly\alpha$ luminosity function which is modeled as a Schechter function, defined as 
\begin{equation}
\label{eq:Schechter}
\phi(L)dL = \phi^*(L/L_*)^\alpha e^{-L/L^*}d(L/L^*)
\end{equation}
where $L$ is the luminosity of the LAEs and $L^*$ is the characteristic LAE luminosity beyond which the power-law form of the function drops off rapidly. Using the 207 LAEs sampled from the  Subaru survey and 58 additional LAE measurements made from \cite{Kawasaki2006}, the best fit Schechter parameters are found to be $\phi^* = 8.5^{+ 3.0}_{- 2.2} \times 10^{-4}$Mpc$^{-3}$ and $L^*_{\rm Ly \alpha} = 4.4^{+0.6}_{-0.6} \times 10^{42}$ergs s$^{-1}$ with fixed $\alpha = -1.5$. The number densities and $Ly\alpha$ luminosity functions are calculated by integrating equation \eqref{eq:Schechter} down to the observed limit of  $L_{\alpha}^{\textrm {min}} = 2.5\times 10^{42}$ ergs/s using the best fit Schechter parameters. The number density is computed to be $\overline{n} = 4.1^{+0.9}_{-0.8} \times 10^{-4}$Mpc$^{-3}$ at $z = 6.6$. The constraints on these quantities include cosmic variance estimates. To mitigate such errors in the future, one can alternatively consider Ly$\alpha$ intensity mapping, which can take advantage of a larger field \citep{Mario}.

The angular correlation function, $\xi(\theta)$, of the 207 measured LAEs are computed using the Landy $\&$ Szalay (LS) estimator. To evaluate the LS estimator, one first creates a pure random catalogue of sources and computes
\begin{equation}
\label{eq:LandySzalay}
\xi(\theta) = [DD(\theta) - 2DR(\theta) + RR(\theta)]/RR(\theta)  ,
\end{equation}
where $DD(\theta)$, $RR(\theta)$ and $DR(\theta)$ are the number of data-data, random-random and data-random pairs normalized by the number of random-random pairs in each angular bin \citep{LS1993}. Observational offsets included in $\xi(\theta)$ due to limited survey area are evaluated by first assuming the true underlying ACF follows a power law of the form $\xi(\theta) = A_\xi \theta^{-\gamma}$, where the values of $A_\xi$  are fit for using the results on $\xi(\theta)$. The observational offset in $\xi(\theta)$ can then be computed using the integral constraint (see \citealt{Ouchi}, \citealt{Peebles1977}). The angular dependence of $\xi(\theta)$ is converted to a dependence on comoving distance $R$ using the Limber approximation \citep{Peebles1980}. The resulting constraints on $\xi(R)$ are quoted in Figure 12 of \cite{Ouchi}. We only use the constraints placed on $\xi(R)$ for comoving scales larger than $> 1.5$Mpc, corresponding to the resolution of our simulation boxes described in Section \ref{sec:SimulationParams}, which corresponding to comoving angular scales larger than $\theta \simeq  35$ arcsec.



\subsection{HERA instrument and Sensitivities}
\label{sec:HERA}
To forecast the constraints that can be placed on $\beta$ using both the Subaru data and a measurement of the 21cm power spectrum, we use HERA as our fiducial 21cm instrument. When completed, HERA will consist of 350 parabolic dishes, each $14\,\textrm{m}$ in diameter with observing frequencies from $50\,\textrm{MHz}$ to $250\,\textrm{MHz}$. Based on its forecasted sensitivities, $> 20\sigma$ detections of the $21\,\textrm{cm}$ power spectrum from the EoR will likely be possible \citep{HERA,AdrianMCMC}. Although we use HERA as our fiducial instrument, the qualitative conclusions presented in Section \ref{sec:MCMCresults} are also valid for other instruments such as the Murchison Widefield Array \citep{Bowman2013,Tingay2013},  the Square Kilometre Array \citep{SKAI}, and the Low Frequency Array \citep{LOFAR2013}.

We model HERA's sensitivities on measurements of $\Delta^2_{\rm{21}}$ using the publicly available code \texttt{21cmSense} \citep{PoberBAOBAB2013,AdrianMCMC}. The code computes the sensitivities on $\Delta^2_{\rm{21}}$ by modeling the instrumental thermal noise using HERA's interferometeric design and survey parameters. Beyond instrumental effects, the code then adds sample variance to the sensitivities. The resulting errors on $\Delta^2_{\rm{21}}$ are computed as 
\begin{equation}
    \varepsilon( { k} ) = D_c^2 Y \frac{{k^3} \xi_{\rm eff} }{2 \pi^2 } \frac{T^2_{\rm sys}}{2 t_{\rm int}} + \varepsilon_{\rm{sample}}
    \label{eq:sensitivities}
\end{equation}
where the first term is the thermal noise specific to HERA and the second term adds sample variance. In the first term,  $T_{\rm{sys}}$ is the antenna temperature of HERA and  $\Omega_{\rm eff}$ is the effective solid angle of the primary beam of each dish \citep{ParsonsLimit2014}. Meanwhile $D_c$ converts angular separations on the sky to comoving distances (see Equation \eqref{eq:X_convert}) and $Y$ converts radial comoving distances $\Delta r_{\parallel}$ to frequency intervals $\Delta \nu$ defined through
\begin{equation}
    Y \equiv \frac{\Delta r_\parallel}{\Delta \nu} = \frac{c}{H_0 \nu_{21}} \frac{(1+z)^2}{E(z)}. 
\end{equation}
where $\nu_{21} \approx 1420\,\textrm{MHz}$ is the rest frequency of the $21\,\textrm{cm}$ line. 
The sample variance is generated using a fiducial EoR inside-out model. Using different fiducial EoR scenarios does not qualitatively change the results in Section \ref{sec:MCMCresults}. 

The 21cm signal is expected to be several orders of magnitude dimmer than the ``foreground" contaminants. The foregrounds are astrophysical in nature and dominate the low-frequency radio spectrum. Fortunately, the foregrounds are spectrally smooth and are expected to lie in the characteristic ``wedge" of cylindrically decomposed Fourier space. The ``wedge" is defined as 
\begin{equation}
\label{eq:wedge}
k_\parallel \leq \left( \frac{D_c}{\nu Y}\right) k_\perp,
\end{equation}
where $\nu$ is the observing frequency, ${k}_\perp$ and $k_\parallel$ are the wavenumbers perpindicular and parallel to the line of sight in cylindrical Fourier space. In generating Equation \eqref{eq:sensitivities}, we adopt the ``moderate" foreground setting in \texttt{21cmSense} which states that modes satisfying Equation \eqref{eq:wedge}, and additional modes up to $0.1\,h\textrm{Mpc}^{-1}$ higher in $k_\parallel$, are contaminated by the foregrounds. This additional ``buffer" accounts for the degree of spectral unsmoothness in the foregrounds which may cause a leakage to higher $k_\parallel$ \citep{21cmsense1}.

\subsection{Markov Chain Monte Carlo Setup for Subaru Constraints}
\label{sec:MCMC_setup_Subaru}
 In this Section we discuss our Markov Chain Monte Carlo (MCMC) setup which we use to place constraints on the EoR parameters. In our MCMC we use only the measurements from the Subaru survey discussion in the previous Section. We place the correlation parameter $\beta$, the EoR parameters, $\zeta$, $M_{\rm {turn}}$, and $R_{\rm {mfp}}$, as well as $f_{\rm{duty}}$ and $M_{\rm{\alpha}}^{\rm min}$ into a single vector $\boldsymbol \theta$. In order to place constraints on $\boldsymbol \theta$, we need to infer the probability of obtaining a particular instance of $\boldsymbol \theta$ given the Subaru dataset $\mathbf{d_{\rm S}}$. This probability distribution, $p(\boldsymbol \theta | \mathbf{d_{\rm S}})$, is the posterior in Bayes' theorem
 \begin{equation}
p(\boldsymbol \theta | \mathbf{d_{\rm S}}) \propto p( \mathbf{d_{\rm S}} | \boldsymbol \theta ) p(\boldsymbol \theta),
\end{equation}
where $p( \mathbf{d_{\rm S}} | \boldsymbol \theta ) $ is the likelihood function and $p(\boldsymbol \theta)$ is our prior. We place uniform priors on all parameters. For the correlation parameter $\beta$, and duty parameter $f_{\rm{duty}}$, we use a uniform prior of  $-1 \leq \beta \leq 1$ and $0 \leq f_{\rm{duty}} \leq 1$ respectively. These ranges encapsulate the entire allowable regions of both their parameter spaces. Values of $\beta$ between $-1 \leq \beta \leq 1$ span the entire range of correlations, while by construction, $f_{\rm{duty}}$ can only have values $0 \leq f_{\rm{duty}} \leq 1$. For $\zeta$, we place the range $10 < \zeta < 100$ which is spans the range of values which are consistent with previous studies such as \cite{kSz}. For $M_{\rm turn}$, we adopt values of $10^7 M_\odot <  M_{\rm turn} <  10^{10} M_\odot$, which are motivated by the atomic cooling threshold and by current constraints on the faint end of UV luminosity functions \citep{Park}. For $R_{\rm mfp}$ we use $3\,\textrm{Mpc} < R_{ \rm mfp} < 80\,\textrm{Mpc}$ which spans the expected range \citep{Rmfp}.  Finally, for  $M_{\rm{\alpha}}^{\rm min}$, we adopt a uniform prior with $10^9 M_\odot \leq M_{\rm{\alpha}}^{\rm min} \leq 10^{11} M_\odot$. These bounds are motivated by \cite{SobacchiAndrei} which find values outside this range to be inconsistent with constraints placed on the $Ly\alpha$ luminosity function by  \cite{MatheeConstraints} and \cite{Ouchi}. 

We use \texttt{21cmFAST} to generate a box of intrinsic LAEs at the redshift of interest $z$. To do this we generate the underlying halo field and then assign an intrinsic LAE luminosity to each virialized halo using Equation \eqref{eq:L_intrinisc}. We resolve all haloes above the virialized mass scale $M = 5 \times 10^8$M$_\odot$. The resulting halos are assigned an intrinsic luminosity according to equation \eqref{eq:L_intrinisc} using model parameters $f_{\rm{duty}}$ and the normalization $M_{\rm{\alpha}}^{\rm min}$. The resulting box of intrinsic LAEs extends $300$Mpc along the line of sight (see Section \ref{sec:SimulationParams} for simulation details). To mimic the the intrinsic LAEs observed by the Subaru Survey which has redshift thickness $\Delta z = 0.1$ (corresponding to $\sim 37$Mpc at $z = 6.6$), we slice our box of intrinsic LAEs into slabs corresponding to a redshift thickness of $\Delta z = 0.1$, in accordance with the redshift uncertainty of the Subaru HSC data.  To compute the likelihood $p( \mathbf{d_S} | \boldsymbol \theta )$, we use \texttt{21cmFAST}  to generate the density and ionization fields for a given set of model parameters  $\zeta$, $R_{\rm mfp}$, and $M_{\rm turn}$. The ionization field is computed from a density field which has been decorrelated with the desired level of $\beta$. For each set of model parameters, we pair the new ionization field to the original box of intrinsic LAEs. We perform separate forecasts using different randomly generated density realizations to account for the cosmic variance. Our conclusions are unchanged for each of these different realizations. The number density $\overline{n}$ and ACF $\xi$ of the model LAEs are computed and compared to the corresponding Subaru measurements of the number densities $\overline{n}_S$, and ACF $\xi_S$, through the likelihood given by
\begin{equation}
\label{eq:bayes_LAE}
    p_{\textrm{LAE}}( \mathbf{d} | \boldsymbol \theta ) \propto \exp \left[-\frac{1}{2} \sum_{R} \frac{ \left(\xi_{\rm model} - \xi_{S} \right)^2}{\varepsilon^2_\xi} \right]   \exp \left[-\frac{1}{2} \frac{ \left(\overline{n}_{\rm model} - \overline{n}_{S} \right)^2}{\varepsilon^2_n} \right],
\end{equation} 
where $\varepsilon_n$ are the Subaru errorbars on the mean LAE number density and $\varepsilon_\xi$ are the errors on the Subaru measurements of the ACF given in \citep{Ouchi}. We symmetrize the errorbars on $\overline{n}$ and $\xi_{\xi}$. In each case we take a conservative approach and symmetrize using the larger error limit. We take measurements of $\overline{n}$ and $\xi(R)$ to be statistically independent which is a reasonable assumption since $\overline{n}$ depends on the mean number of LAEs while $\xi$ depends only on the overdensity $\delta_n$, which is mean zero. We approximate the $R$ bins in $\xi(R)$ as being statistically independent.
\subsection{Markov Chain Monte Carlo Setup for Joint Subaru $\&$ 21cm Forecasts}
\label{sec:MCMC_setup_Joint}
In this Section we build on the setup from the previous Section and explore the constraints that can be placed on $\beta$ using measurement of the 21cm power spectrum $\Delta^2_{\rm 21}$ in addition to the Subaru measurements described in Section \ref{sec:Subaru_Constraints}. If we assume these probes are independent from one another, the posterior for joint measurements between LAE and 21cm probes can be written using Bayes theorem as,
 \begin{equation}
 \label{eq:bayes_joint}
p(\boldsymbol \theta | \mathbf{d_{\rm S}  ,  \Delta^2_{21}}) \propto p_{\rm 21}( \mathbf{\Delta^2_{21}} | \boldsymbol \theta ) p_{\textrm{LAE}}( \mathbf{d_{\rm S}} | \boldsymbol \theta ) p(\boldsymbol \theta),
\end{equation}
where $p_{\textrm{LAE}}( \mathbf{d} | \boldsymbol \theta )$ is the likelihood function for the Subaru measurements (discussed in the previous Section),  $p_{\textrm{21}}( \mathbf{d} | \boldsymbol \theta )$ is the likelihood for measurements of the 21cm power spectrum and $p(\theta)$ is the prior on our parameters. Our priors are identical to those in Section \ref{sec:MCMC_setup_Subaru}.

To evaluate the likelihood $p_{\textrm{21}}( \mathbf{d} | \boldsymbol \theta )$, we generate model predictions for the density, ionization fields and temperature fields from 21cmFAST simulations for a given set of model parameters $\mathbf{\theta}$. To account for the decorrelation between the ionization and density fields, we regenerate the ionization field for this set of EoR parameters, but with the desired level of decorrelation from the original ionization field as specified by the $\beta$ parameter. This updated ionization field is then used with the original density field to form a brightness temperature field using Equation \eqref{eq:dTb}. The power spectrum $\Delta^2_{\rm model}$ of $\delta T_b$ is computed using Equation \eqref{eq:power_spectrum}. We compare the power spectrum $\Delta^2_{\rm model}$ of the model temperature field to the fiducial power spectrum $\Delta^2_{21} (k,z)$ using the likelihood, 
\begin{equation}
    p_{\rm 21}( \mathbf{d} | \boldsymbol \theta ) \propto \exp \left[-\frac{1}{2} \sum_{z , k} \frac{ \left(\Delta^2_{\rm model} - \Delta^2_{21} \right)^2}{\varepsilon^2} \right],
\end{equation} 
where we have assumed that all the $k$ and $z$ bins are statistically independent. We consider redshifts $ 7.5 \le z \le 8.5$ in steps of $\Delta z = 0.5$, corresponding to observational bandwidth $\Delta \nu \equiv \nu_{21} \Delta z/(1+z)^2$ of each redshift bin. We choose these redshift ranges because they correspond to one of HERA's relatively clean observation windows. We exclude bins $ k > 0.75\,\textrm{Mpc}^{-1}$ for computational simplicity as the HERA error bars are large in that regime and inclusion of larger $k$ bins do not add alter our forecasts significantly. In this forecast, we use a mock HERA observation of the power spectrum generated using a fiducial set of EoR parameter values, $\zeta_0 = 25$, $M_{\textrm{turn}, 0} = 5\times 10^8 M_{\odot}$, $R_{\textrm{mfp},0} = 30\,\textrm{Mpc}$, and $\beta_0 = 0.936$ unless otherwise indicated. To sample our posterior distribution, we use the affine invariant MCMC package \texttt{emcee} \citep{emcee}.  




\section{Results}
\label{sec:MCMCresults}

In this section we present the results of our MCMCs and discuss their implications. We separate the results into two sections. In Section \ref{sec:Subaru_Constraints}, we place constraints on the EoR parameters using existing measurements of the number density, and ACF, of LAEs at redshift $z = 6.6$ from the Subaru experiment, discussed in Section \ref{sec:subaru}. In Section \ref{sec:Subaru_HERA}, we forecast the types on constraints that we can place on the EoR parameters (including $\beta$), using measurements of the 21cm power spectrum and $\overline{n}$, $\xi$, of LAEs. For this multiple probe forecast we use Subaru measurements of the number density and ACF of LAEs at redshift $z = 6.6$ and a HERA measurement of the 21cm power spectrum between redshifts  $z = 7.5$ to $z = 8.5$.

\subsection{Subaru Constraints}
\label{sec:Subaru_Constraints}
 \begin{figure}
  \includegraphics[width=0.49\textwidth]{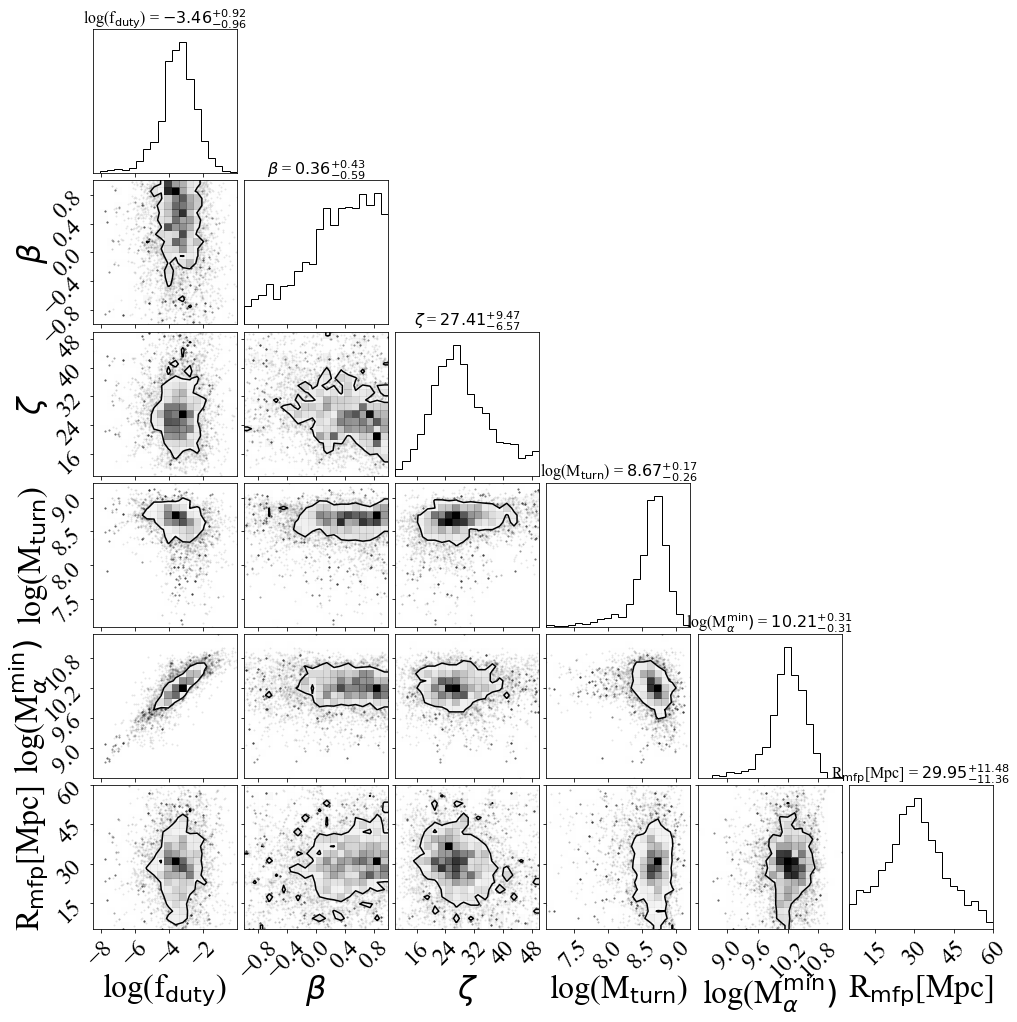}
  \caption{Posterior distribution for the EoR and LAE parameters using the Subaru data described in Section \ref{sec:MCMC_setup_Subaru}. Displayed are the $68 \%$ CR contours. Since the $68\%$ CR contours include both the $\beta > 0$, and $\beta < 0$ regions of parameter space, measurement of $\overline{n}$ and $\xi$ at $z = 6.6$ cannot constrain the sign of $\beta$ to within $68\%$ credibility.}
  \label{fig:mcmc_lae}
\end{figure}
In Figure \ref{fig:mcmc_lae} we show the posterior of our MCMC analysis using the Subaru measurements of $\overline{n}$, and $\xi$, at $z = 6.6$. We find an evident degeneracy between $M_{\rm{\alpha}}^{\rm {min}}$ and $f_{\rm {duty}}$. To see why, consider an intrinsic LAE field with LAE detection threshold $M_{\rm{\alpha}}^{\rm {min}}$ corresponding to the minimum luminosity detectable by our experiment, $L_{\rm {\alpha}}^{\rm{min}}$. Physically, $M_{\rm{\alpha}}^{\rm {min}}$ is the halo mass that corresponds to the minimum luminosity $L_{\rm {\alpha}}^{\rm{min}}$ (the faintest luminosity detectable by our instruments). Larger values of $M_{\rm{\alpha}}^{\rm {min}}$ decrease the number of intrinsic LAEs that would be detectable by our instruments. This in turn decreases the measured number density of LAEs. In order to fit to the number density $\overline{n} = 4.1^{+0.9}_{-0.8}\times 10^{-4}$ by Subaru, this scenario requires increasing $f_{\rm {duty}}$ which leads to the observed degeneracy. Values of $M_{\rm{\alpha}}^{\rm {min}}$ larger than $M_{\rm{\alpha}}^{\rm {min}} > 10^{11} M_\odot$  require $f_{\rm {duty}} > 1$ in order to be consistent with the already constrained value of $\overline{n}$. This is not possible and so models which require values $M_{\rm{\alpha}}^{\rm {min}} > 10^{11} M_\odot$, have already been ruled out \citep{SobacchiAndrei}. 
We do not find an explicit degeneracy between $\beta$ and the other EoR parameters. Ruling out an outside-in driven reionization is tantamount to placing constraints on the sign of $\beta$. However we see from the 1D $\beta$ posterior in Figure \ref{fig:mcmc_lae} that the $68 \%$ credibility region (CR) of $\beta$ is not entirely contained within the $\beta > 0$ region, suggesting that measurements of $\overline{n}$ and $\xi$ at $z = 6.6$ alone is insufficient to rule out uncorrelated, or outside-in scenarios. In Figure \ref{fig:mcmc_lae} we see that all $\beta$ models are broadly consistent with measurements of $\xi$ from Subaru. We see that our models can't distinguish between the extreme inside-out models. The Universe is significantly ionized at redshift $z = 6.6$, and so the number density and ACF of LAEs is not very sensitive to changes in $\beta$. As a result, the Subaru data is unable to concretely rule out uncorrelated and outside-in reionization scenarios at $68\%$CR. At higher redshift, there is increased sensitivity of $\overline{n}$ and $\xi$ to $\beta$. Measurements of $\overline{n}$ and $\xi$ at higher redshifts are required in order to place tighter constraints on the EoR morphology using only LAE data. 

From the posterior, we see that $68\%$ of the 1D marginalized posterior of $M_{\rm{turn}}$ lie within the range $\textrm{log} M_{\rm{turn}} = 8.67^{+0.17}_{-0.26}$. Therefore we find that using the Subaru data, we can place constraints on the order of magnitude of  $M_{\rm{turn}}$ at $68\%$ CR, while  $\zeta$ is constrained within $\zeta = 27.41^{+9.47}_{-6.57}$ at $68\%$ CR. 





\subsection{Joint Subaru $\&$ HERA Forecasts}
\label{sec:Subaru_HERA}
 \begin{figure}
  \includegraphics[width=0.49\textwidth]{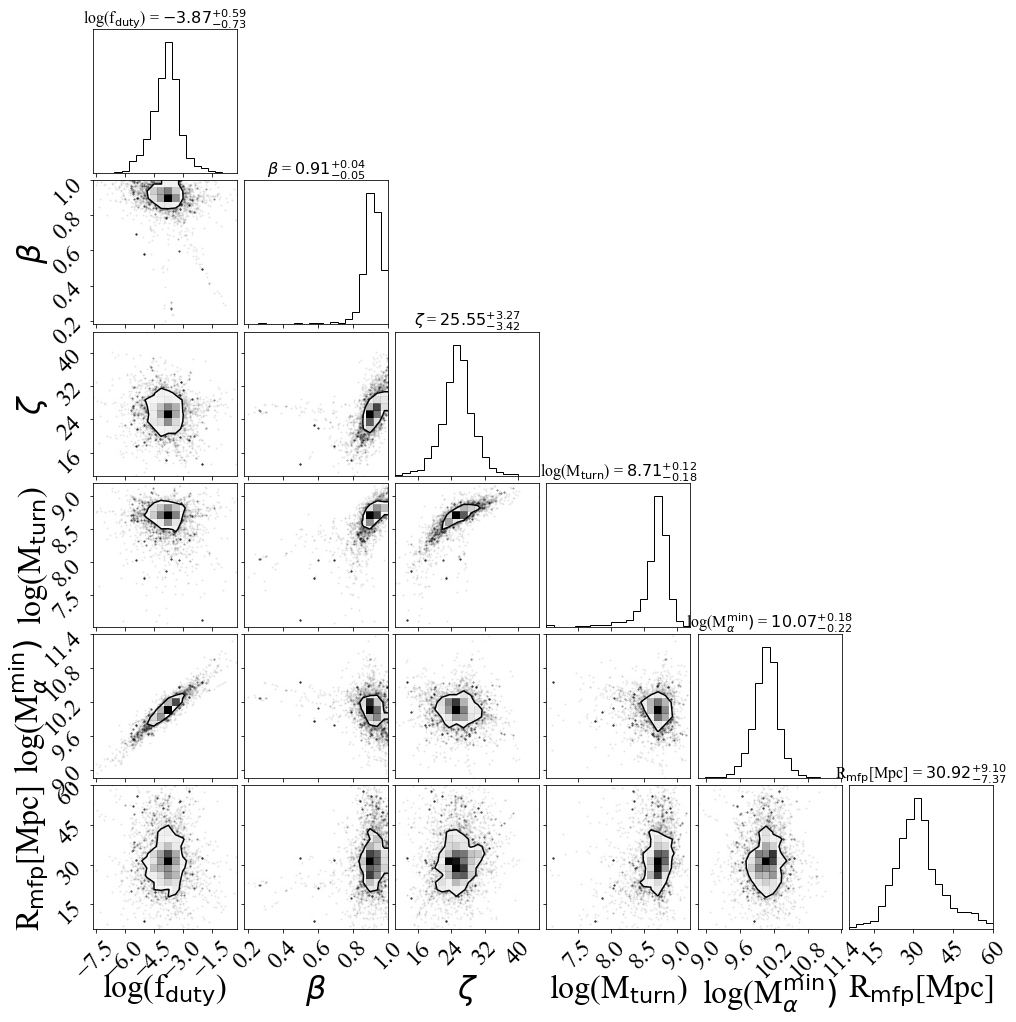}
  \caption{Posterior distribution for the EoR and LAE parameters using the Subaru data at $z = 6.6$ and the 21cm power spectrum from $7.5 \le z \le 8.5$. The  $68 \%$ CR are entirely contained within $\beta > 0 $ suggesting that if reionization proceeds as inside-out, measurement of $\overline{n}$ and $\xi$ of LAE as well as $\Delta^2_{\rm 21}$ can rule out uncorrelated and outside-in reionization with $68\%$ credibility.}
    \label{fig:mcmc_joint}
\end{figure}


In this scenario we forecast the type of constraints that can be placed on the EoR morphology using joint measurements of the LAE number density and ACF, along with a HERA measurement of the 21cm power spectrum between redshifts $z = 7.5$ to $z = 8.5$.  These redshifts are chosen to maximize the greatest signal to noise ratio for $\beta$ in $\Delta^2_{\rm{21}}$ after taking the instrument sensitivities of $\Delta^2_{\rm{21}}$ into account. Since these 21cm observations do not yet exist, we use a fiducial inside-out reionization model with fiducial parameters $\zeta_0 = 25$ , $M_{\rm{turn},0} = 5\times 10^8$M$_\odot$, $R_{\rm{mfp},0} = 30$Mpc and $\beta_0 = 0.936$. This fiducial reionization scenario is consistent with the constraints placed on these parameters using the Subaru measurements of $\overline{n}$ and $\xi$ in Section \ref{fig:mcmc_lae}. 

The results of this forecast are shown in Figure \ref{fig:mcmc_joint}. Our interpretation of the degeneracies between $f_{\rm{duty}}$ and $M_{\alpha}$ are identical to Section \ref{sec:Subaru_Constraints}. We see that adding information from $\Delta^2_{\rm{21}}$ significantly improves our ability to discern between EoR models. From the posterior of this measurement in Figure \ref{fig:mcmc_joint}, we see that the $99\%$ credibility region of $\beta$ lie entirely within $\beta > 0$. This is the predominantly inside-out region of $\beta$ parameter space. Therefore measurements of $\overline{n}$ and $\xi$ of LAEs at $z = 6.6$, and a HERA measurement of $\Delta^2_{\rm{21}}$ at $7.5 \le z \le 8.5$ can rule out uncorrelated and outside-in reionization scenarios with $99\%$ credibility. An identical forecast performed in \cite{me!} using the same $k$ bins, redshift range, and fiducial reionization model, but without the LAE data, was previously studied. In this forecast it was found that $\beta$ models in the range $0.9 \le \beta \le 1$, i.e. extreme inside-out scenarios, were equally likely. This suggests that measurement of the 21cm power from Section \ref{sec:MCMC_setup_Joint} alone is not able to distinguish between extreme inside-out $\beta$ models. From the $\beta$ contours in Figure \ref{fig:mcmc_joint}, we see that the inclusion of Subaru data allows us to distinguish between these extreme inside-out scenarios.

\section{Conclusion}
\label{sec:Conclusion}

The correlation between density and ionization fields is crucial to our understanding of reionization. Many probes of the EoR are sensitive to this underlying correlation. LAEs are one such probe because Ly$\alpha$ photons are sensitive to the coupling of the ionized regions with respect to the underlying intrinsic LAEs. In this paper we explore how the statistics of LAEs are affected by the morphology of the EoR. To do this we introduce a parameter $\beta$, which parameterizes the correlation between density and ionization fields. Under this parametrization we study how the correlation between density and ionziation fields affects the number density and clustering of LAEs. We find that changing $\beta$ results in altering the number density of LAEs. Outside-in driven reionization scenarios ($\beta < 0$), decrease the mean number density of LAEs compared to inside-out driven scenarios ($\beta > 0$). We also find that varying $\beta$ affects the apparent clustering of LAEs. At higher redshifts, outside-in driven reionization scenarios produce an increase in the clustering signal of LAEs compared to that of inside-out. 

Using Subaru measurements of the clustering and number density of LAEs at $z = 6.6$, we place constraints on $\beta$. We find that measurements of these quantities alone cannot rule out uncorrelated scenarios at $68\%$ credibility. The ACF and mean number density of LAEs have reduced sensitivity to $\beta$ late in reionization where most of the IGM is ionized. Upcoming HERA limits of the 21cm power spectrum will also allow us to place constraints on $\beta$. We forecast the type of constraints that we can place on $\beta$ using both HERA measurements of $\Delta^2_{\rm{21}}$ at the midpoint of reionization, and Subaru measurements of the LAE number density and ACF at $z = 6.6$. We find that including Subaru measurements of $\overline{n}$ and $\xi$ at $z = 6.6$ can improve the constraints placed on $\beta$ using only HERA measurements of $\Delta^2_{\rm{21}}$. The LAE information at $z = 6.6$ can help distinguish between extreme inside-out scenario and further rule out uncorrelated scenarios to levels over $99\%$ credibility. These results show that LAE information at $z = 6.6$, i.e. when reionization is expected to be nearing its end, are already enough to help distinguish between models with different density ionization correlations. As we observe LAEs at higher redshifts, combining these probes will have even greater potential, and will help shed light on one of the most crucial properties of reionization.

\section*{Acknowledgements}

The authors are delighted to acknowledge helpful discussions with Anne Hutter, Jordan Mirocha and Hannah Fronenberg.  We acknowledge support from the New Frontiers in Research Fund Exploration grant program, a Natural Sciences and Engineering Research Council of Canada (NSERC) Discovery Grant and a Discovery Launch Supplement, the Sloan Research Fellowship, the William Dawson Scholarship at McGill, as well as the Canadian Institute for Advanced Research (CIFAR) Azrieli Global Scholars program. This research was enabled in part by support provided by Calcul Quebec (\url{www.calculquebec.ca}), WestGrid (\url{www.westgrid.ca}) and Compute Canada (\url{www.computecanada.ca}).



\section*{Data Availability}
The software code underlying this article will be shared on reasonable request to the corresponding author.

\bibliographystyle{mnras}
\bibliography{example} 




\appendix


\bsp	
\label{lastpage}
\end{document}